\documentclass[11pt]{article}
\usepackage{graphicx}
\usepackage{longtable}
\makeatletter
 
  \@addtoreset{equation}{section}
\makeatother
\def\beqar {\begin{eqnarray}}
\def\eeqar {\end{eqnarray}}
\def\beq {\begin{equation}}
\def\eeq {\end{equation}}
\def\A{{\cal A}}
\def\B{{\cal B}}
\def\C{{\cal C}}

\def\S{{\cal S}}

\def\al{\alpha}
\def\bt{\beta}
\def\del{\delta}

\def\ga{\gamma}

\def\ka{\kappa}
\def\ep{\epsilon}
\def\la{\lambda}
\def\La{\Lambda}
\def\om{\omega}
\def\Om{\Omega}
\def\th{\theta}

\def\si{\sigma}

\def\d{\partial}
\def\bd{{\bar \partial}}

\def\ba{{\bar a}}
\def\bz{{\bar z}}

\def\bom{{\bar \omega}}

\def\hf{\frac{1}{2}}

\def\<{\langle}\def\bra{\langle}
\def\>{\rangle}\def\ket{\rangle}
\def\Li{{\rm Li}}
\def\re{{\rm Re}}
\def\im{{\rm Im}}
\def\Tr{{\rm Tr}}

\def\Path{{\rm P}}
\def\cp{{\bf CP}}

\begin{document}

\begin{titlepage}
\null\vspace{-62pt} \pagestyle{empty}
\begin{center}
% \rightline{CCNY-HEP-/10}
\vspace{0.8truein}

{\Large\bf
Application of abelian holonomy formalism \\
\vspace{.35cm}
to the elementary theory of numbers} \\

%%%%%%%%%%%%%%%%%%%%%%%%%%%%%%%%%%%%%%%%%%%%%%%%%%%%%%%%%%
\vspace{0.8in} {\sc Yasuhiro Abe} \\
\vskip .12in {\it Cereja Technology Co., Ltd.\\
1-13-14 Mukai-Bldg. 3F, Sekiguchi \\
Bunkyo-ku, Tokyo 112-0014, Japan } \\
\vskip .07in {\tt abe@cereja.co.jp}\\
\vspace{0.8in}
%%%%%%%%%%%%%%%%%%%%%%%%%%%%%%%%%%%%%%%%%%%%%%%%%%%%%%%%%%%%
\centerline{\large\bf Abstract}
\end{center}

\noindent
We consider an abelian holonomy operator in two-dimensional
conformal field theory with zero-mode contributions.
The analysis is made possible by use of a geometric-quantization
scheme for abelian Chern-Simons theory on $S^1 \times S^1 \times {\bf R}$.
We find that a purely zero-mode part of the holonomy operator
can be expressed in terms of Riemann's zeta function.
We also show that a generalization of linking numbers can be
obtained in terms of the vacuum expectation values of
the zero-mode holonomy operators.
Inspired by mathematical analogies between linking numbers
and Legendre symbols, we then apply these results to a
space of ${\bf F}_p = {\bf Z}/ p {\bf Z}$ where $p$ is an odd prime number.
This enables us to calculate ``scattering amplitudes'' of identical
odd primes in the holonomy formalism.
In this framework, the Riemann hypothesis can be interpreted
by means of a physically obvious fact, {\it i.e.}, there is no
notion of ``scattering'' for a single-particle system.
Abelian gauge theories described by the zero-mode holonomy operators
will be useful for studies on quantum aspects
of topology and number theory.

\end{titlepage}
%%%%%%%%%%%%%%%%%%%%%%%%%%%%%%%%%%%%%%%%%%%%%%%%%%%%%%%%%%%%%
\pagestyle{plain} \setcounter{page}{2} %\baselineskip =14pt

%\tableofcontents
%\vskip 1.2cm

%%%%%%%%%%%%%%%%%%%%%%%%%%%%%%%%%%%%%%%%%%%%%%%%%%%%%%%%%%%%%%%%%%
\section{Introduction}

The concept of holonomy in conformal field theory or, more precisely,
holonomy of what is called the Knizhnik-Zamolodchikov (KZ) connection in mathematics
has been developed in search of a geometric understanding of
the conformal field theory \cite{Kohno:2002bk}.
This geometric approach has an advantage in obtaining topological invariants.
Particularly, it is known that the holonomy is useful in construction of the so-called Witten
invariant which is originally formulated as a partition
function of three-dimensional Chern-Simons theory \cite{Witten:1988hf}.
The holonomy in conformal field theory is therefore closely related to the knot theory.
In fact, the holonomy can be defined as a linear representation of a braid group.
Since knots, and links in general, can be generated by the braid group,
the holonomy is indeed an indispensable element in application of
(conformal) field theory to topology.

Recently, use of such a holonomy operator in a form that is more familiar to physicists
is proposed as a non-perturbative formulation of non-abelian gauge theories \cite{Abe:2009kn}.
In this approach, which we call holonomy formalism, tree-level scattering amplitudes of gluons
are shown to be generated by a holonomy operator that is defined in a $\cp^1$ fiber of twistor space.
Notice that twistor space $\cp^{3}$ can be regarded as a $\cp^1$-bundle over compact spacetime $S^4$;
thus twistor space (or its supersymmetric version) is naturally required in order to
incorporate four-dimensional spacetime information.
An essential structure of the holonomy formalism is, however, encoded in
the holonomy operator of the $\cp^1$ fiber on which a
two-dimensional conformal field theory is defined.
This formulation is developed along the line of the so-called spinor-momenta
or spinor-helicity formalism.
(For details of this formalism, see \cite{Abe:2009kn} and references therein.)
In order to further understand the validity and usefulness of
the holonomy formalism and its essentials, it is therefore
natural to investigate an abelian holonomy operator
in conformal field theory without any spacetime or twistor-space information.
This is a main motivation of the present paper.

The abelian holonomy operator that we discuss here serves as a topological ``skeleton'' of
abelian gauge theories, including, in particular, Maxwell's theory of electromagnetism.
Thus topological invariants that can be generated by the abelian holonomy should include
Gauss's linking number.
It is well-known that the linking numbers can be expressed in terms of a partition
function of abelian Chern-Simons theory (see, for example,
\cite{Polychronakos:1989-90}-\cite{Guadagnini:2010gy}).
The liking numbers can also be expressed as degrees of mapping, or winding numbers, of a map
$\phi :  S^1 \times S^1 \longrightarrow S^2$.
(For the notion of winding number, see, {\it e.g.}, \cite{Fulton:1995bk}.)
In fact, the linking numbers naturally arise
from an abelian Chern-Simons theory on $S^1 \times S^1 \times {\bf R}$
which can be interpreted as a Wess-Zumino-Witten (WZW) model on torus.
As we review in section 3, this can neatly be shown by use of geometric quantization for
the toric abelian Chern-Simons theory \cite{BosNair:1989,NairBook,Abe:2008wn}.
We shall apply this geometric scheme to the abelian holonomy formalism such that
generalization of linking numbers can be obtained in terms of the abelian holonomy operator.
Generalization and application of linking numbers in different physical contexts
can also be found in \cite{Horowitz:1989km}-\cite{GarciaCompean:2009zh}.

In the present paper, we try to convey a novel perspective on the linking numbers
by applying the abelian holonomy formalism to the elementary theory of numbers.
This is motivated by the following mathematical results.
Interestingly, the concept of linking number can be applied to
odd primes and, by use of Galois groups, one can define a linking number (mod 2) of
$p$ and $q$, denoted by $lk (p,q)$, where $p$ and $q$ are two distinct odd primes.
This is shown by Morishita \cite{Morishita:2009gt}.
We do not discuss details of $lk (p,q)$ here but the upshot of
Morishita's result is that $lk (p, q)$ is
related to the Legendre symbol by %\cite{Morishita:2009gt}
\beq
    ( -1 )^{lk (q, p)} = \left( \frac{q^*}{p} \right)
    \label{1-1}
\eeq
where $q^{*} = (-1)^{\frac{q-1}{2}} q$.
The Legendre symbol $\left( \frac{q^*}{p} \right)$ is defined by
\beq
    \left( \frac{q^*}{p} \right) = \la_p (q^*) =
    \left\{
    \begin{array}{l}
    + 1 ~~~ \mbox{if $q^*$ is a quadratic residue modulo $p$} \, ; \\
    - 1 ~~~ \mbox{otherwise} \, .  \\
    \end{array}
    \right.
    \label{1-2}
\eeq
That $q^*$ is a quadratic residue modulo $p$ means that there exists
an integer $n$ such that $n^2 \equiv q^*$ (mod $p$).
In the elementary theory of numbers, the Legendre symbol appears in the Gauss sum
\beq
    \widehat{\la}_p = \sum_{x = 1}^{p-1} \la_p (x) \,  e^{i \frac{2\pi}{p} x} \, .
    \label{1-3}
\eeq
Notice that the sum is taken over the elements of ${\bf F}_{p}^{\times} = {\bf F}_p - \{ 0 \}$
where ${\bf F}_p = {\bf Z}/ p {\bf Z}$.
Since the Gauss sum takes a value of complex number, the Legendre symbol gives a map
\beq
    \la_p : ~ {\bf F}_{p}^{\times} \, \rightarrow \, {\bf C} \, .
    \label{1-4}
\eeq
This suggests that we can interpret $\widehat{\la}_p$
as a Fourier transform of $\la_p (x)$ in a space of mod $p$.
(For description of the Gauss sum as a Fourier transform,
see, {\it e.g.}, \cite{Ono:1987bk}.)

At first glance, the relations (\ref{1-1})-(\ref{1-4}) seem irrelevant
to the algebraic construction of holonomies.
However, as mentioned earlier, the holonomy formalism is closely related to
knot theory and linking numbers can be generated by an abelian holonomy operator.
Thus, the facts that the linking number is essentially equivalent
to the Legendre symbol and that the Gauss sum can be interpreted as a Fourier transform of
the Legendre symbol in a space of finite field ${\bf F}_{p}^{\times}$
are quite suggestive.
Namely, application of the holonomy formalism
to elementary number theory will be useful to obtain
something analogous to scattering amplitudes of prime numbers.
In other words, it is presumably possible to
express a generating function for such ``scattering amplitudes'' of primes
in terms of the abelian holonomy operator suitably defined in ${\bf F}_{p}^{\times}$.
A main goal of the present paper is to give an explicit realization of
such a generating function.
Roughly speaking, this means that our main goal is to
obtain a field theoretic description of the idea
that there is a correspondence between knots and primes, {\it i.e.},
\beq
    ( \mbox{integer} ) ~ = ~ \prod_{i} ( \mbox{prime} )_{i}
    ~ \longleftrightarrow ~ \prod_{i} ( \mbox{knot} )_{i} ~ = ~ ( \mbox{link} ) \, ,
    \label{1-5}
\eeq
in the context of holonomy formalism.
For mathematical studies on the relation between knots and primes,
one may refer to \cite{Kohno:2006bk}.

Another intriguing aspect of the abelian holonomy formalism which we would like
to deliver in the present paper is that, if we include zero-mode contributions and consider
only the zero-mode part of the abelian holonomy operator,
then the zero-mode holonomy operator can be expressed in
terms of Riemann's zeta function.
It is well-known that Riemann's zeta function can be
represented by an iterated integral \cite{Kohno:2002bk}.
In section 4, we shall show that the zero-mode part of the abelian holonomy operator
contains essentially the same iterated-integral representation as Riemann's zeta function.
The fact that the abelian holonomy operator is represented by an iterated integral
is not surprising if we remember that the WZW model can be
defined by an iterative integral \cite{Efraty:1992gk}.
(For the use of an iterative integral in physics, see also \cite{Zupnik:1987vm}.)

Riemann's zeta function is intimately related to prime numbers
by the formula of Euler's product.
In applying the zero-mode holonomy operator to a space of ${\bf F}_{p}^{\times}$,
we shall make a detailed argument that one can interpret the Gauss sum as
an operator which is relevant to creation of odd primes.
Along with the discussion on the generating function for
``scattering amplitudes'' of primes, we then further investigate
the zero-mode holonomy operator in relation to quantum realization of
Riemann's zeta function.
We propose that clarification of this relation will lead to a new, if speculative,
physical interpretation of the Riemann hypothesis.
(For recent studies on related topics, see, {\it e.g.},
\cite{Sierra:2008se}-\cite{Giri:2009pk}.)

The paper is organized as follows.
In the next section, we briefly review the holonomy formalism,
following \cite{Abe:2009kn}, and present an explicit
definition of the abelian holonomy operator.
In section 3, we review the geometric-quantization scheme for
the toric $U(1)$ Chern-Simons theory, following \cite{Abe:2008wn}.
We shall show that linking numbers naturally arise from a holomorphic wavefunction
of the toric $U(1)$ Chern-Simos theory.
In section 4, utilizing the results in the previous sections, we construct and calculate
an abelian holonomy operator with zero-mode contributions.
We show that generalization of linking numbers can be obtained from
such a holonomy operator.
We also show that the zero-mode part of the abelian holonomy operator
can be expressed in terms of Riemann's zeta function by
use of an iterated-integral representation.
In section 5, we apply the zero-mode holonomy operator
to a space of ${\bf F}_{p}^{\times}$ and discuss how one can obtain
a generating function for ``scattering amplitudes'' of prime numbers.
We shall also  propose a new interpretation of the Riemann hypothesis in
the context of abelian holonomy formalism.
Lastly, we shall present some concluding remarks.

%%%%%%%%%%%%%%%%%%%%%%%%%%%%%%%%%%%%%%%%%%%%%%%%%
\section{Review of abelian holonomy operators}

In this section we review essentials of the holonomy formalism,
following the construction introduced in \cite{Abe:2009kn}.
The holonomy formalism is originally developed in explaining multi-gluon
amplitudes in helicity-based calculations.
Basic physical operators are then given by creation
operators of gluons with helicity $\pm$, which can be identified as
ladder operators of the $SL(2 ,{\bf C})$ algebra;
these are denoted by $a_{i}^{(\pm)}$ in (\ref{2-1}).
An abelian version of the formalism is thus obtained simply
by replacing a role of gluons with that of photons.
As mentioned earlier, in this paper we shall focus on holonomies of
two-dimensional conformal field theories, rather than
dealing with twistor space to construct four-dimensional theories.
Therefore, strictly speaking, the notion of helicity and even that of 4-momentum
are not appropriate.
The use of $SL(2 , {\bf C})$ algebra is however persistent in either case.
Thus we can still use $a_{i}^{(\pm)}$ as physical operators in two dimensions. Here
the sign $\pm$ of $a_{i}^{(\pm)}$ no longer corresponds to helicity but some analog of it.

\vskip 0.5cm \noindent
\underline{Braid groups, KZ equation and comprehensive gauge fields}
%\underline{Emergence of Iwahori-Hecke algebra}

Bearing in mind the above notes on helicity, we now consider
a multi-photon system in the holonomy formalism.
The Hilbert space of the system is given by
$V^{\otimes n} = V_1 \otimes V_2 \otimes \cdots \otimes V_n$
where $V_ i$ $(i=1,2,\cdots,n)$ denotes a Fock space that creation operators
of the $i$-th particle with helicity $\pm$ act on.
Such physical operators $a_{i}^{(\pm)}$ are
given by ladder operators that form a part of the $SL(2, {\bf C})$ algebra.
The algebra can be expressed as
\beq
    [ a_{i}^{(+)}, a_{j}^{(-)}] = 2 a_{i}^{(0)} \, \del_{ij}  \, , ~~
    [ a_{i}^{(0)}, a_{j}^{(+)}] = a_{i}^{(+)} \, \del_{ij} \, , ~~
    [ a_{i}^{(0)}, a_{j}^{(-)}] = - a_{i}^{(-)} \, \del_{ij}
\label{2-1}
\eeq
where Kronecker's deltas show that the non-zero commutators are obtained only when $i = j$.
The remaining of commutators, those expressed otherwise, all vanish.

The physical configuration space of $n$ photons
is given by $\C = {\bf C}^n / \S_n$, where $\S_n$ is the rank-$n$ symmetric group.
The $\S_n$ arises from the fact that photons are bosons.
The complex number ${\bf C}$ corresponds to a local coordinate of
$\cp^1$ on which, in the context of spinor-momenta formalism, a spinor momentum
of the $i$-th photon is defined.
In this paper we do not strictly follow the spinor-momenta formalism.
Thus it is not appropriate to use a momentum-representation of $a_{i}^{(\pm)}$.
However, for a purpose of obtaining abelian holonomy operators of conformal field theory,
we can still label these operators in terms of complex coordinates of $\cp^1$
because it is this $\cp^1$ where a conformal field theory is defined.
In the next section, we shall deform the topology of $\cp^1 = S^2$ to
that of torus $T^2 = S^1 \times S^1$ and shall consider zero-mode contributions
to the abelian holonomies.

It is well-known that the fundamental homotopy group of $\C = {\bf C}^n / \S_n$
is given by the braid group, $\Pi_1 (\C) = \B_n$.
The braid group $\B_n$ has generators, $b_1 , b_2 , \cdots ,
b_{n-1}$, and they satisfy the following relations:
\beq
    \begin{array}{cc}
    b_i b_{i+1} b_i = b_{i+1} b_i b_{i+1} & ~~~ \mbox{if $~ |i - j| = 1$} \, , \\
    b_i b_j = b_j b_i  & ~~~ \mbox{if $~ |i - j| > 1$} \\
    \end{array}
    \label{2-2}
\eeq
where we identify $b_n$ with $b_1$.
To be mathematically rigorous, the ${\bf C}$ of $\C = {\bf C}^n / \S_n$ should
be replaced by $\cp^1$ which is represented by the local coordinate $z$.
Since ${\bf C}$ can be obtained from $\cp^1$ by excluding points at infinity,
the replacement can be done with ease.
At the level of braid generators, this can be
carried out by imposing the following relation \cite{Chari:1994pz}
\beq
    (b_1 b_2 \cdots b_{n-2} b_{n-1} ) (b_{n-1} b_{n-2} \cdots b_2 b_1) = 1 \, .
    \label{2-3}
\eeq
The braid group that satisfies this condition on top of (\ref{2-2}) is
called a {\it sphere} braid group $\B_n (\cp^1)$, while the previous one is called
a {\it pure} braid group $\B_n ({\bf C}) = \B_n $.
Thus, bearing in mind this subsidiary condition, we can identify $\C = {\bf C}^n / \S_n$
as the physical configuration space of interest.

Now, mathematically, a linear representation of a braid gruop
is equivalent to a monodromy representation of the
Knizhnik-Zamolodchikov (KZ) equation.
The KZ equation is an equation that a function on $\C$ satisfies in
general. We can denote such a function as $\Psi (z_1 , z_2 , \cdots , z_n)$,
where $z_i$ ($i = 1, 2, \cdots, n$) represents the local complex coordinate of $\cp^1$.
The KZ equation is then expressed as
\beq
    \frac{\d \Psi }{ \d z_i} =  \frac{1}{\kappa}
    \sum_{~ j \, (j \ne i)} \frac{\Om_{ij} \Psi}{z_i - z_j}
    \label{2-4}
\eeq
where $\kappa$ is a non-zero constant called the KZ parameter.
We now introduce logarithmic differential one-forms
\beq
    \om_{ij} = d \log (z_i - z_j) = \frac{ d z_i - d z_j}{z_i - z_j} \, .
    \label{2-5}
\eeq
Notice that these satisfy the identity
\beq
    \om_{ij} \wedge \om_{jk} + \om_{jk} \wedge \om_{ik} + \om_{ik} \wedge \om_{ij} = 0
    \label{2-6}
\eeq
where the indices are ordered as $i < j < k$.
The quantity $\Om_{ij}$ in the KZ equation is a bialgebraic operator.
In terms of the operators of $SL(2, {\bf C})$ algebra in (\ref{2-1}), this can be
defined as
\beq
    \Om_{ij} = a_{i}^{(+)} \otimes a_{j}^{(-)} + a_{i}^{(-)} \otimes a_{j}^{(+)}
    + 2 a_{i}^{(0)} \otimes a_{j}^{(0)} \, .
    \label{2-7}
\eeq
In the case of $i = j$, this becomes the quadratic Casimir of $SL(2, {\bf C})$ algebra
which acts on the $i$-th Fock space $V_i$.
Action of $\Om_{ij}$ on $V^{\otimes n} = V_1 \otimes V_2 \otimes \cdots \otimes V_n$
can be written as
\beq
    \sum_{\mu} 1 \otimes \cdots \otimes 1 \otimes \rho_i (I_{\mu})
    \otimes 1 \otimes \cdots \otimes 1 \otimes \rho_j (I_{\mu}) \otimes 1 \otimes \cdots
    \otimes 1
    \label{2-7-1}
\eeq
where $I_\mu$ ($\mu = 0,1,2$) are elements of the $SL(2, {\bf C})$ algebra
and $\rho$ denotes its representation.
Introducing the one-form
\beq
    \Om =  \frac{1}{\kappa} \sum_{1 \le i < j \le n} \Om_{ij} \, \om_{ij} \, ,
    \label{2-8}
\eeq
we can then rewrite the KZ equation (\ref{2-4}) as a differential equation
\beq
    D \Psi = (d - \Om) \Psi = 0
    \label{2-9}
\eeq
where $D = d - \Om$ can be regarded as a covariant exterior derivative.

From the explicit form of (\ref{2-7}), one can show
\beqar
    \label{2-10}
    [ \Om_{ij} , \Om_{kl} ] &=& 0  ~~~~~ \mbox{($i,j,k,l$ are distinct),} \\  \label{2-11}
    [ \Om_{ij} + \Om_{jk} , \Om_{ik} ] &=& 0  ~~~~~ \mbox{($i,j,k$ are distinct).}
\eeqar
In mathematical literature, these are called the infinitesimal braid relations.
Remarkably, by use of these relations along with (\ref{2-6}),
the flatness of $\Om$, {\it i.e.}, $d \Om - \Om \wedge \Om = 0$, can be shown.
(For a proof of this, see \cite{Kohno:2002bk,Abe:2009kn}.)
Therefore, it is possible to define a holonomy of $\Om$, which
gives a general linear representation of a braid group on
the Hilbert space $V^{\otimes n}$. This is the monodromy representation
of the KZ equation. The Hilbert space $V^{\otimes n}$
can then be identified as the space of conformal blocks for the KZ equation.

We now consider applications of these mathematical results to
the multi-photon system.
As mentioned before, the physical operators of photons are given by $a_{i}^{(\pm)}$.
The operator $\Om_{ij}$ may not be appropriate to describe photons since
its action on the Hilbert space is represented by (\ref{2-7-1}), which involves $a_{i}^{(0)}$.
We need to modify $\Om_{ij}$'s such that the operators $a_{i}^{(0)}$
are treated somewhat unphysically.
We then introduce a following ``comprehensive'' gauge one-form
\beq
    A =  \frac{1}{\kappa} \sum_{1 \le i < j \le n} A_{ij} \, \om_{ij}
    \label{2-12}
\eeq
where $A_{ij}$ is defined as a bialgebraic operator
\beq
    A_{ij} = a_{i}^{(+)} \otimes a_{j}^{(0)} + a_{i}^{(-)} \otimes a_{j}^{(0)} \, .
    \label{2-13}
\eeq
Notice that, from the explicit form of $A_{ij}$, we can also show that
the bialgebraic quantity $A_{ij}$ satisfy the relations (\ref{2-10}) and (\ref{2-11}).
(For details of this fact, see \cite{Abe:2009kn}.)
These relations are the only conditions
for the flatness or integrability of $A$.
Thus, as in the case of $\Om$, we can also obtain the expression
\beq
    DA = dA - A \wedge A = - A \wedge A = 0
    \label{2-14}
\eeq
where $D$ is now a covariant exterior derivative $D = d - A$.
This relation guarantees the existence of holonomies for the comprehensive
gauge field $A$.

Although the bialgebraic structures of  $\Om$ and $A$ are
different, the constituents of these remain the same, {\it i.e.}, they are given
by $a_{i}^{(0)}$ and $a_{i}^{(\pm)}$. Thus, we can use the same Hilbert space $V^{\otimes n}$
and physical configuration $\C$ for both $\Om$ and $A$.
The KZ equation of $A$ is then given by $D \Psi = (d - A) \Psi = 0$, where
$\Psi$ is a function of a set of complex variables $(z_1 , z_2 , \cdots z_n)$.
This suggests that the inverse of the KZ parameter $\kappa$ can be interpreted
as a coupling constant of the gauge theory.

\vskip 0.5cm \noindent
\underline{Abelian holonomy operators}

A holonomy of $A$ can be given by a general solution to the KZ equation
$D \Psi = (d - A) \Psi = 0$.
The construction is therefore similar to that of Wilson loop operators.
In the present formalism, rank-$n$ differential manifolds
are physically relevant for the construction.
Thus, we need differential $n$-forms
in terms of $A$ in order to define an appropriate holonomy operator.
Further, an analog of Wilson loop should be defined on $\C$.
These requirements lead to the following definition of the holonomy operator.
\beq
    \Theta_{R, \ga} (z) = \Tr_{R, \ga} \, \Path \exp \left[
    \sum_{r \ge 2} \oint_{\ga} \underbrace{A \wedge A \wedge \cdots \wedge A}_{r}
    \right]
    \label{2-15}
\eeq
where $\ga$ represents a closed path on $\C$ along which the integral is
evaluated and $R$ denotes the representation of a gauge group.
For the abelian group $U(1)$, the representation can be trivially determined.
In the following, we explain the meanings of the symbol $\Path$ and the trace $\Tr_{R, \ga}$,
respectively.

The symbol $\Path$ denotes a ``path ordering'' of the numbering indices $i$, $j$.
The meaning of this symbol can be understood as follows.
The exponent in (\ref{2-15}) can be expanded as
\beq
    \sum_{r \ge 2}^{\infty} \oint_{\ga} \underbrace{A \wedge \cdots \wedge A}_{r}
    = \sum_{r \ge 2}^{\infty} \oint_{\ga}  \sum_{ (i<j) } A_{i_1 j_1} A_{i_2 j_2} \cdots A_{i_r j_r}
    \bigwedge_{k=1}^{r} \om_{i_k j_k}
    \label{2-16}
\eeq
where $(i < j)$ means that the set of indices $(i_1, j_1, \cdots , i_r , j_r)$
are ordered such that $1 \le i_1 < j_1 \le r$,$\cdots$,$1 \le i_r < j_r \le r$.
Notice that the coupling constant $1/ \ka$ is absorbed in $A$; we shall use this convention
unless otherwise mentioned in the following.
Notice also that, if we have $r = 1$, we can define (\ref{2-16}),
or the exponent of (\ref{2-15}), as zero.
For $r=1$, we cannot define either $A$ defined in (\ref{2-12}) or
$\om_{ij}$ in (\ref{2-5}) but it is natural to consider $\om_{ij}$ vanishing
since in this case the quantity $( z_i - z_j )$ can be treated as a fixed variable.
(We shall come back to this definition later in the end of section 5.)
The symbol $\Path$ means that the numbering indices are further received ordering conditions
$1 \le i_1 < i_2 < \cdots < i_r \le r$ and $2 \le j_1 < j_2 < \cdots < j_r \le r+1$
where $r+1$ is to be identified with $1$. This automatically leads to
the following expression
\beq
    \Path \sum_{r \ge 2}^{\infty} \oint_{\ga} \underbrace{A \wedge \cdots \wedge A}_{r}
    = \sum_{r \ge 2}^{\infty} \oint_{\ga}  A_{1 2} A_{2 3} \cdots A_{r 1}
    \, \om_{12} \wedge \om_{23} \wedge \cdots \wedge \om_{r 1}
    \label{2-17}
\eeq
which shows that the basis of the holonomy operators (\ref{2-15}) is given by $\om_{i \, i+1}$
under the ``path ordering'' operation for the numbering indices.

Now the commutator $[A_{12}, A_{23}]$ can be calculated as
\beqar
    \nonumber
    [A_{12}, A_{23}]
    &=& a_{1}^{(+)} \otimes a_{2}^{(+)} \otimes a_{3}^{(0)}
    - a_{1}^{(+)} \otimes a_{2}^{(-)} \otimes a_{3}^{(0)}  \\
    && \!\!\! + \, a_{1}^{(-)} \otimes a_{2}^{(+)} \otimes a_{3}^{(0)}
    - a_{1}^{(-)} \otimes a_{2}^{(-)} \otimes a_{3}^{(0)} \, .
    \label{2-18}
\eeqar
Equation (\ref{2-17}) is then written as
\beqar
    \nonumber
    \Path \sum_{r \ge 2}^{\infty} \oint_{\ga} \underbrace{A \wedge \cdots \wedge A}_{r}
    &=& \sum_{r \ge 2}^{\infty} \oint_{\ga}  A_{1 2} A_{2 3} \cdots A_{r 1}
    \, \om_{12} \wedge \om_{23} \wedge \cdots \wedge \om_{r 1} \\
    \nonumber
    &=& \sum_{r \ge 2}^{\infty} \frac{1}{2^{r+1}} \sum_{(h_1, h_2, \cdots , h_r)}
    (-1)^{h_1 + h_2 + \cdots + h_r} \\
    && ~~~ \times \,
    a_{1}^{(h_1)} \otimes a_{2}^{(h_2)} \otimes \cdots \otimes a_{r}^{(h_r)}
    \, \oint_{\ga} \om_{12} \wedge \cdots \wedge \om_{r1}
    \nonumber \\
    \label{2-19}
\eeqar
where $h_{i}$ denotes $h_{i} = \pm = \pm 1$ ($i=1,2,\cdots, r$).
In the above expression, we define
$a_{1}^{(\pm)} \otimes a_{2}^{(h_2)} \otimes
\cdots \otimes a_{r}^{(h_r)} \otimes a_{1}^{(0)}$ as
\beqar
    \nonumber
    a_{1}^{(\pm)} \otimes a_{2}^{(h_2)} \otimes \cdots \otimes a_{r}^{(h_r)} \otimes a_{1}^{(0)}
    & \equiv &
    \hf [a_{1}^{(0)} , a_{1}^{(\pm)}] \otimes a_{2}^{(h_2)} \otimes \cdots \otimes a_{r}^{(h_r)} \\
    &=&
    \pm \hf a_{1}^{(\pm)} \otimes a_{2}^{(h_2)} \otimes \cdots \otimes a_{r}^{(h_r)}
    \label{2-20}
\eeqar
where we implicitly use an antisymmetric property for the indices $(1,2, \cdots , r)$
as indicated in (\ref{2-16}) or (\ref{2-17}).

The trace $\Tr_{R, \ga}$ in the definition (\ref{2-15}) means
traces over the generators of a $U(1)$ group and a braid group.
The former trace is trivial and the latter, which is called a braid trace,
can be realized by a sum over permutations of the numbering indices.
Thus the trace $\Tr_{R, \ga}$ over the exponent of (\ref{2-15}) can be
expressed as
\beqar
    &&  \Tr_{\ga} \Path \sum_{r \ge 2}^{\infty} \oint_{\ga}
    \underbrace{A \wedge \cdots \wedge A}_{r} \nonumber \\
    &=& \sum_{r \ge 2}
    \sum_{\si \in \S_{r-1}} \oint_{\ga}  A_{1 \si_2} A_{\si_2 \si_3} \cdots A_{\si_r 1}
    \, \om_{1 \si_2} \wedge \om_{\si_2 \si_3} \wedge \cdots \wedge \om_{\si_r 1}
    \label{2-21}
\eeqar
where the summation of $\S_{r-1}$ is taken over the
permutations of the elements $\{2,3, \cdots, r \}$,
with the permutations labeled by
$\si=\left(%
\begin{array}{c}
  2 \, 3 \cdots r \\
  \si_2 \si_3 \cdots \si_r \\
\end{array}%
\right)$.

From the expression (\ref{2-19}), we find that,
with a suitable normalization for
$\oint_\ga \om_{12} \wedge \cdots \wedge \om_{r1}$, the holonomy operator
can be used as a generating function for all physical states of photons
on the Hilbert space $V^{\otimes n}$.
The holonomy operator (\ref{2-15}) is therefore a very universal operator.
As a summary of this section, we recapitulate the definition of
the abelian holonomy operator $\Theta_{R, \ga} (z)$ below.
\beqar
    \Theta_{R, \ga} (z) &=& \Tr_{R, \ga} \, \Path \exp \left[
    \sum_{r \ge 2} \oint_{\ga} \underbrace{A \wedge A \wedge \cdots \wedge A}_{r}
    \right]
    \label{2-22} \\
    A &=&  \frac{1}{\kappa} \sum_{1 \le i < j \le n} A_{ij} \, \om_{ij}
    \label{2-23} \\
    \om_{ij} & = & d \log (z_i - z_j ) = \frac{ d z_i - d z_j }{ z_i - z_j }
    \label{2-24} \\
    A_{ij} &=& a_{i}^{(+)} \otimes a_{j}^{(0)} + a_{i}^{(-)} \otimes a_{j}^{(0)}
    \label{2-25}
\eeqar

%%%%%%%%%%%%%%%%%%%%%%%%%%%%%%%%%%%%%%%%%%%%%%%%%
%\section{Zero-mode contributions}
\section{Geometric quantization and linking numbers}

In the following sections, we shall consider zero-mode contributions to
the abelian holonomy operator
$\Theta_{R, \ga} (z)$ by changing the topology of $\cp^1 = S^2$ to $T^2 = S^1 \times S^1$.
This means that we impose a double periodicity condition on
the complex variables $z_i$ ($i = 1,2, \cdots , n$):
\beq
    z_i \rightarrow z_i + m_i + n_i \tau
    \label{3-1}
\eeq
where $m_i$ and $n_i$ are integers and $\tau = \re \tau + i \im \tau$ is
the modular parameter of the torus.
We shall carry out this analysis by use of a geometric-quantization scheme
developed in \cite{Abe:2008wn}
for the $U(1)$ Chern-Simons theory on $S^1 \times S^1 \times {\bf R}$.
In this preparatory section, we briefly review this scheme and show
its relation to linking numbers.

\vskip 0.5cm \noindent
\underline{Holonomies of torus}

Torus can be described in terms of two real coordinates $\xi_1$, $\xi_2$ with
periodicity of $\xi_r \rightarrow \xi_r + m$ ($r=1,2$) where $m$ is any integer.
In other words, $\xi_r$ take real values in $0 \le \xi_r \le 1$,
with the boundary values $0$, $1$ being identical.
Complex coordinates of torus can be
parametrized as $z = \xi_1 + \tau \xi_2$ where $\tau \in {\bf C}$ is
the modular parameter of the torus.
This parametrization is equivalent to (\ref{3-1}).
Notice that we can absorb the real part of $\tau$ into $\xi_1$
without losing generality.
In the following, we then assume $\re \tau = 0$, {\it i.e.},
\beq
    \tau = \re \tau + i \im \tau = i \im \tau \, .
    \label{3-2}
\eeq

There are two noncontractible cycles on torus, conventionally labeled
as $\al$ and $\bt$ cycles.
Holonomies of torus can be defined along these cycles as
\beq
    \int_\al \om = 1 \, , ~~~ \int_\bt \om = \tau
    \label{3-3}
\eeq
where the holomorphic one-form $\om = \om(z) dz$ is
a zero mode of the anti-holomorphic derivative $\d_\bz = \frac{\d}{\d \bz}$.
The normalization of $\om$ and its complex conjugate $\bom$ is given by
\beq
    \int dz d\bz \, \bom \wedge \om  \, = \, i 2 \, \im \tau \, .
    \label{3-4}
\eeq
In constructing a theory on torus, we need to take account of
factors that involve $\om$ and $\bom$.

Let $a \in {\bf C}$ be a complex physical variable of the zero mode.
As we discuss in a moment, an abelian gauge potential on torus
can be parametrized solely by this complex variable $a$ and its conjugate.
Notice that these also satisfy the double periodicity
\beq
    a \rightarrow a + m + n \tau \, , ~~~ \ba \rightarrow \ba + m + n \bar{\tau}
    \label{3-5}
\eeq
where $m$ and $n$ are integers which correspond to
the winding numbers of $\al$ and $\bt$ cycles, respectively.

Using the assumption (\ref{3-2}), we can rewrite
the holonomies (\ref{3-3}) in a more convenient form:
\beqar
    \oint_{\al_r} \om_s &=&  (\im \tau) \, \ep_{rs} \, ,
    \label{3-6} \\
    \om_1 &=& (d \bz - dz ) / 2i  \, = \, -\im \tau d\xi_2  \, ,
    \label{3-7} \\
    \om_2 &=& (\tau d\bz - \bar{\tau} dz) / 2i \, = \,  \im \tau d\xi_1
    \label{3-8}
\eeqar
where $\ep_{rs}$ ($r,s = 1,2$) denotes the rank-2 Levi-Civita symbol
and $\al_1$, $\al_2$ correspond to the $\al$ and $\bt$ cycles, respectively.
Notice that we can set $\om (z) = 1$ in (\ref{3-3}) by identifying
the $\al$ and $\bt$ cycles with loop integrals along
the variables $\xi_1$ and $\xi_2$, respectively.
Normalization for $\om_1$ and $\om_2$ is given by
\beq
    \int dz d\bz \, \frac{\om_1}{\im \tau} \wedge \frac{\om_2}{\im \tau} \, = \, 1 \, .
    \label{3-9}
\eeq

We now reparametrize the complex physical variables as
\beq
    a_1 = \ba - a \, , ~~~ a_2 = \tau \ba - \bar{\tau} a \, .
    \label{3-10}
\eeq
These are compatible with the choice of $\om_1$ and $\om_2$.
Under the transformations of (\ref{3-5}), $a_1$ and $a_2$ vary as
\beq
    \del a_1 \rightarrow (-2i \im \tau) n \, , ~~~
    \del a_2 \rightarrow (2 i \im \tau) m  \, .
    \label{3-11}
\eeq
From (\ref{3-6}) and (\ref{3-11}), we find
\beq
    \exp \left(
    {\oint_{\al_2} \frac{\pi \om_1}{\im \tau} \frac{\del a_2}{\im \tau}}
    \right)
    = e^{-i 2 \pi m}  \, , ~~
    \exp \left(
    {\oint_{\al_1} \frac{\pi \om_2}{\im \tau} \frac{\del a_1}{\im \tau}}
    \right)
    = e^{-i 2 \pi n} \, .
    \label{3-12}
\eeq

\vskip 0.5cm \noindent
\underline{Geometric quantization of the toric $U(1)$ Chern-Simons theory}

We now consider geometric quantization of the $U(1)$ Chern-Simons theory,
following the line of \cite{BosNair:1989,NairBook} in a slightly different manner.

In the temporal gauge, Chern-Simons gauge potentials
can be described by the spatial components.
In terms of the real coordinates $\xi_r$ $(r= 1,2)$ on torus,
these can be parametrized as
\beqar
    A_{\xi_1} &=& i \d_{\xi_1} \th ~ + ~ \frac{\pi \om_2}{\im \tau} \frac{a_1}{\im \tau} \, ,
    \nonumber \\
    A_{\xi_2} &=& i \d_{\xi_2} \th ~ + ~ \frac{\pi \om_1}{\im \tau} \frac{a_2}{\im \tau}
    \label{3-13}
\eeqar
where $\d_{\xi_r}$ denotes $\frac{\d}{\d \xi_r}$ and
$\th = \th ( \xi_1 , \xi_2 )$ is a function of $\xi_1$, $\xi_2$.
From (\ref{3-12}), we can easily find that
holonomies of $A_{\xi_r}$ are invariant under
the transformations of (\ref{3-5}).
Notice also that, under a certain gauge where $\th$ is a constant,
the gauge potentials are parametrized purely by zero modes.
(For non-abelian theories, however, it is not possible to express
$A_{\xi_1}$, $A_{\xi_2}$ entirely by zero modes; see \cite{Abe:2008wn} for details.)
Thus, in terms of the complexified variables, the gauge potentials
of the $U(1)$ Chern-Simons theory on $S^1 \times S^1 \times {\bf R}$
can be parametrized by
\beq
    A_z = \frac{ \pi \om}{\im \tau} \ba \, , ~~~
    A_\bz  = \frac{ \pi \bom }{\im \tau} a \, .
    \label{3-14}
\eeq

In the program of geometric quantization, a ``holomorphic'' wavefunction
$\Psi [ A_\bz ]$ of the $U(1)$ Chern-Simons theory generally satisfies
the so-called polarization condition
\beq
    \left( \d_a + \frac{1}{2} \d_a K \right) \, \Psi [ A_\bz ] \, = \, 0
    \label{3-15}
\eeq
where $K$ is a K\"{a}hler potential that is associated with
the phase space of Chern-Simons theory in the $A_0 = 0$ gauge.
The condition (\ref{3-15}) leads to the specific form
\beq
    \Psi[ A_\bz ] = e^{-\frac{K}{2}} \psi[ A_\bz ]
    \label{3-16}
\eeq
where $\psi [ A_\bz ]$ is a purely (anti)holomorphic function of $A_\bz$.
In the present case, physical variables are given by $a$ and $\ba$.
Thus we can express the holomorphic wavefunction (\ref{3-16}) as
\beq
    \Psi[ A_\bz] \equiv \Psi[a] = e^{-\frac{K (a,\ba)}{2}} f(a)
    \label{3-17}
\eeq
where $f (a)$ is a function of $a$.
This is a function defined on torus. Thus it is natural
to require the invariance of $f (a)$ under the transformation
$a \rightarrow a + m + n \tau$.
In the following, we shall show this relation by a suitable
choice of the K\"{a}hler potential $K ( a, \ba )$.

We first notice that there is an ambiguity in defining
K\"{a}hler potentials. What is essential in physics in a geometric analysis is
the K\"{a}hler form to start with.
There are a number of K\"{a}hler potentials that leads to the same  K\"{a}hler form.
In the present case, form (\ref{3-4}) and (\ref{3-14}), we can define
the K\"{a}hler form of zero modes as
\beq
    \Om^{(\tau )} ~=~ \frac{k}{2 \pi} da \wedge d\ba \int_{z,\bz}
    \left( \frac{ \pi \bom}{\im \tau}  \right)
    \wedge \left( \frac{ \pi \om}{\im \tau}  \right)
    ~=~ i \frac{\pi k}{\im \tau} da \wedge d \ba
    \label{3-18}
\eeq
where the integral is taken over $dz d\bz$ and
$k$ is the level number associated to the abelian Chern-Simons theory.
The corresponding K\"{a}hler potentials can generally be expressed as
\beq
    W(a, \ba) =  \frac{\pi k}{\im \tau} a \ba + g(a) + h (\ba)
    \label{3-19}
\eeq
where $g(a)$ and $h (\ba)$ are arbitrary functions of $a$ and $\ba$, respectively.
Thus there are infinite choices for the K\"{a}hler potentials.
We need to choose an appropriate one depending on specific purposes.
This is what we are going to do now.

From our normalization (\ref{3-4}) and (\ref{3-9}), we find a relation
\beq
    da \wedge d \ba \int_{z,\bz} \bom \wedge \om \, = \,
    da_1 \wedge d a_2 \int_{z,\bz} \frac{\om_2}{\im \tau }\wedge \frac{\om_1}{\im \tau}
    \label{3-20}
\eeq
where we use
\beq
    \om = \frac{\om_2 - \tau \om_1}{\im \tau}  \, , ~~
    \bom = \frac{\om_2 - {\bar \tau} \om_1}{\im \tau} \,.
    \label{3-21}
\eeq
In terms of $a_1$ and $a_2$, the zero-mode K\"{a}hler form
can then be expressed as
\beqar
    \Om^{(\tau )} &=& \!\! \frac{k}{2 \pi}
    \left( \frac{\pi}{\im \tau} \right)^2 da \wedge d \ba
    \int_{z,\bz} \bom \wedge \om \, = \, \frac{k}{2 \pi}
    \left( \frac{\pi}{\im \tau} \right)^2 (2i \im \tau) da \wedge d \ba
    \nonumber \\
    &=& \!\! \frac{k}{2 \pi}
    \left( \frac{\pi}{\im \tau} \right)^2 da_1 \wedge d a_2
    \int_{z,\bz} \frac{\om_2}{\im \tau} \wedge \frac{\om_1}{\im \tau}
     \, = \, - \frac{k}{2 \pi}
    \left( \frac{\pi}{\im \tau} \right)^2  da_1 \wedge d a_2 \, .
    \nonumber \\
    \label{3-22}
\eeqar
A K\"{a}hler potential corresponding to the second line in (\ref{3-22})
can be written as
\beq
    K (a,\ba) = \frac{i \pi k}{2(\im \tau)^2}
    (\ba -a)(\tau \ba - \bar{\tau} a ) \, .
    \label{3-23}
\eeq
We shall choose this $K(a,\ba)$ as our K\"{a}hler potential for zero modes.

The symplectic potential corresponding to $\Om^{(\tau )}$ can be expressed as
\beqar
    \A^{(\tau )} &=&  \frac{\pi k}{4 (\im \tau)^2} \int_{z,\bz}
    \left(
    \frac{\om_2 a_1}{\im \tau} \wedge \frac{\om_1}{\im \tau} da_2
    - \frac{\om_1 a_2}{\im \tau} \wedge \frac{\om_2}{\im \tau} da_1
    \right) \nonumber \\
    &=& - \frac{\pi k}{4 (\im \tau)^2}
    ( a_1 da_2 + a_2 da_1) \, .
    \label{3-24}
\eeqar
Notice that we can interpret $\A^{(\tau )}$ as a $U(1)$ gauge potential on torus.
In this sense, the transformations (\ref{3-5}) serve as gauge transformations.
From (\ref{3-11}) we find that a variation of $\A^{(\tau )}$ under $a \rightarrow a + m + n \tau$
is given by
\beqar
    \A^{(\tau )} & \rightarrow & \A^{(\tau )} \, + \, d \La_{m,n}
    \label{3-25} \\
    \La_{m,n} &=& - i \frac{\pi k}{2 \im \tau} \, ( m a_1 - n a_2 ) \, .
    \label{3-26}
\eeqar
Gauge invariance of the wavefunction $\Psi[a]$ in (\ref{3-17})
is then realized by imposing
\beq
    e^{i \La_{m,n}} \Psi[a] \, = \, \Psi[a + m + n \tau] \, .
    \label{3-27}
\eeq
This leads to the following relation
\beq
    f(a) \, = \, e^{- i \pi k mn} f(a + m + n \tau) \, .
    \label{3-28}
\eeq
Therefore the holomorphic function $f (a)$ is
invariant under $a \rightarrow a + m + n \tau$,
given that the level number $k$ is quantized by
even integers, {\it i.e.},
\beq
    k \, \in  \, 2 {\bf Z} \, .
    \label{3-29}
\eeq
This level-number quantization condition is well-known for
the toric $U(1)$ Chern-Simons theory.
Here we have reviewed this fact in the context of geometric quantization.

An inner product of the holomorphic wavefunctions can be expressed as
\beq
    \< 1 | 2 \> \, = \,
    \int d\mu(a,\ba) \, e^{-K(a,\ba)} \, \overline{f_{1}(a)} f_2 (a)
    \label{3-30}
\eeq
where $d \mu (a , \ba )= [ da d \ba ]$ denotes the integration measure of
the zero-mode variables up to normalization and
$\overline{f_{1}(a)}$ is the complex conjugate of the function $f_{1} (a)$.
Action of the derivative $\frac{\d}{\d a}$ on $f(a)$ leads to
a factor of $\frac{\pi k}{\im \tau} \ba$.
This is a natural consequence of geometric quantization with
our K\"{a}hler form (\ref{3-18}) that implies the ``phase space'' of the zero-mode
variables is given by $\left( \frac{\pi k}{\im \tau} \ba ,  a \right)$.
Thus, regardless the choices of K\"{a}hler potentials, we always have the operative
relation:
\beq
    \frac{\pi k}{\im \tau} \ba \, \leftrightarrow  \frac{\d}{\d a} \, .
    \label{3-31}
\eeq
Notice also that the holomorphic wavefunction $\Psi [a]$ and its inner product (\ref{3-30})
completely specify the toric $U(1)$ Chern-Simons theory
as a topological quantum field theory.

\vskip 0.5cm \noindent
\underline{Realization of linking numbers}

The level-number quantization (\ref{3-29}) is derived form
the invariance of the holomorphic function $f (a)$ under $a \rightarrow a + m + n \tau$.
In quantum theory, however, physical observables are given by the square of wavefunctions.
Thus, in the present case, $| f (a) |^2$ is physically more interesting
and, in this sense, we can relax the condition (\ref{3-29}).
This allows us to set the level number $(k \in {\bf N})$ to
\beq
    k = 1 \, .
    \label{3-32}
\eeq
If $k > 1$, we can absorb it into one of the winding numbers, say $m$.
The ``phase factor'' $\exp ( i \pi k m n )$ in (\ref{3-28}) then becomes
\beq
    ( - 1 )^{mn} \equiv ( - 1 )^{lk ( \al^m , \bt^n )}
    \label{3-33}
\eeq
where we introduce the notation $lk ( \al^m , \bt^n )$
because the value $mn$, as seen in a moment,
can naturally be interpreted as a linking number
of $\al$ and $\bt$ cycles; remember that $m$ and $n$ respectively denote the
winding numbers of these cycles on torus.

%%%%%%%%%%%%%%%%%%%%%%%%% figure %%%%%%%%%%%%%%%%%%%%%%%%%
\begin{figure} [htbp]
\begin{center}
\includegraphics[width=9.5cm]{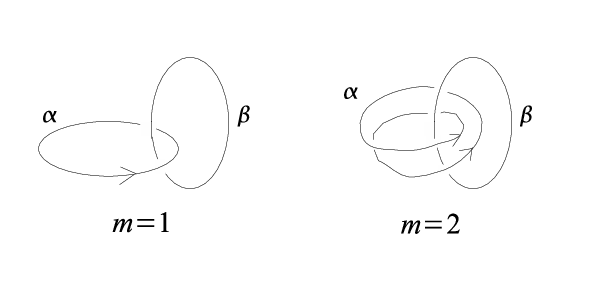}
\caption{How the $\al$ and $\bt$ cycles are entangled with each other in the cases of
linking number $m = 1, 2$. The arrows correspond to the (positive) sings of $m$.}
\label{figNum01}
\end{center}
\end{figure}
%%%%%%%%%%%%%%%%%%%%%%%%% figure %%%%%%%%%%%%%%%%%%%%%%%%%

In defining the linking number, we can in fact fix one integer, say $n$,
because we can always unravel one cycle so that the entanglement of two cycles
can be measured by a winding number of one cycle around the other.
We therefore fix the value of $n$ to identity
\beq
    n = 1 \, .
    \label{3-34}
\eeq
Indeed, as seen in Figure \ref{figNum01}, the linking number of $\al$ and $\bt$ cycles
can be defined by $m = lk ( \al^m , \bt )$.
The relation (\ref{3-28}) is then reduces to
\beq
    f (a) \, = \, (-1)^m   f (a + m + i \im \tau ) \, .
    \label{3-35}
\eeq

\section{Zero-mode contributions to the abelian holonomy operator}
%\vskip 0.5cm \noindent
%\underline{Zero-mode contributions to the holonomy operator}

\vskip 0.5cm \noindent
\underline{Construction of $\Theta_{R, \ga} (z,a)$}

We now apply the results in section 3 to the abelian holonomy operator
$\Theta_{R, \ga} (z)$ which has been defined in (\ref{2-22}).
{\it Our strategy is to incorporate multiple zero-mode variables
$a_i$ $(i= 1,2, \cdots , n)$ into $\Theta_{R, \ga} (z)$ by use of an abelian gauge theory.}
Thus we first consider the covariantization of the comprehensive
``photon'' field $A$:
\beq
    A \, \rightarrow \, \widetilde{A} \, = \, A \, + \, A^{(\tau )}
    \label{4-1}
\eeq
where $A^{(\tau)}$ is an analog of $A$ for zero modes.
We can then define $A^{(\tau)}$ as
\beq
    A^{(\tau)} = \sum_{i < j} A^{(\tau)}_{ij} \om_{ij}
    \label{4-2}
\eeq
where $\om_{ij}$ are the logarithmic one-forms given in (\ref{2-5}) and
$A^{(\tau)}_{ij}$ is a bialgebraic operator for zero modes.
In the following, we shall determine possible forms of $A^{(\tau)}_{ij}$.

As discussed in (\ref{3-14}), we can parametrize a holomorphic gauge field
of zero modes as
\beq
    A_{z} = \frac{\pi \ba }{\im \tau} \om \, .
    \label{4-3}
\eeq
From (\ref{3-31}) we know that the factor $\frac{\pi \ba}{\im \tau}$,
acting on functions of $a$, can be replaced by a derivative operator $\d_a$.
Notice that the derivative operators do not have to obey the $SL(2, {\bf C})$ algebra.
This is a crucial difference from the photon-creation operators $a_{i}^{(\pm)}$.
In fact, by introducing a conjugate operator $\bd_a$ which corresponds
to $\ba$, we find that these operators obey the following bosonic algebra:
\beq
    [ \d_{a_i} , \bd_{a_j} ] = \del_{ij} \, , ~~
    [ \d_{a_i} , \d_{a_j} ] = [ \bd_{a_i} , \bd_{a_j} ] = 0 \, .
    \label{4-4}
\eeq
An analog of a number operator can be defined by
\beq
    N_i = \bd_{a_i} \d_{a_i}
    \label{4-5}
\eeq
which satisfies
\beq
    [ N_i , \d_{a_j} ] = - \del_{ij} \d_{a_i}  \, , ~~
    [ N_i , \bd_{a_j} ] = \bd_{a_i} \del_{ij} \, .
    \label{4-6}
\eeq
In the context of geometric quantization, $N_i$ corresponds to the area of
the $i$-th phase space $\left( \frac{\pi \ba_i }{\im \tau} , a_i \right)$.
In this sense, we can also consider the bosonic algebra (\ref{4-4})
as an analog of the Heisenberg algebra.
As in the previous case, from this algebra
it is natural for us to assume the form of $A^{(\tau)}_{ij}$ as
\beq
    A^{(\tau)}_{ij} =  N_i \otimes \d_{a_j} + N_i \otimes \bd_{a_j} \, .
    \label{4-7}
\eeq

In defining the holonomy operator for $\widetilde{A}$, all what we need
is to show the infinitesimal braid relations, {\it i.e.},
\beqar
    [ \widetilde{A}_{ij} , \widetilde{A}_{kl} ] &=& 0
    ~~~ \mbox{($i,j,k,l$ are distinct)}
    \label{4-8} \\
    \left[  \widetilde{A}_{ij} + \widetilde{A}_{jk}  ,  \widetilde{A}_{ik}  \right] &=& 0
    ~~~ \mbox{($i,j,k$ are distinct)}
    \label{4-9}
\eeqar
As in the case of $A$, the first relation is obvious since any
commutators of two operators that have distinct numbering indices vanish.
Thus it is sufficient to show the second relation
and this can easily be checked as follows.
We first expand $[ \widetilde{A}_{ij} , \widetilde{A}_{ik} ]$ as
\beqar
    [ \widetilde{A}_{ij} , \widetilde{A}_{ik} ] &=&
    [ A_{ij} + A_{ij}^{(\tau)} , A_{ik} + A_{ik}^{(\tau)} ]
    \nonumber \\
    &=& [A_{ij} , A_{ik} ]
    + [A_{ij} , A_{ik}^{(\tau)} ] + [A_{ij}^{(\tau)} , A_{ik} ]
    + [A_{ij}^{(\tau)} , A_{ik}^{(\tau)} ]
    \label{4-10}
\eeqar
Each of these terms can be evaluated by using the definition of a commutator
for bialgebraic operators
\beqar
\nonumber
[ c_i \otimes d_i , c_j \otimes d_j ] &=& [c_i , c_j ] \otimes d_i \otimes d_j \\ \nonumber
&& \!\!\!\! + \, c_i \otimes [d_i , c_j ] \otimes d_j \\
&& \!\!\!\! + \, c_i \otimes c_j \otimes [d_i , d_j]
\label{4-11}
\eeqar
where $c_i$ and $d_i$ $(i = 1,2,\cdots , n)$ denote arbitrary operators;
these are usual algebraic operators.
As shown in \cite{Abe:2009kn}, the first term vanishes:
\beqar
    \nonumber
    [A_{ij} , A_{ik} ] &=& [a_{i}^{(+)} \otimes a_{j}^{(0)} , a_{i}^{(-)} \otimes a_{k}^{(0)} ]
    + [a_{i}^{(-)} \otimes a_{j}^{(0)} , a_{i}^{(+)} \otimes a_{k}^{(0)} ] \\
    &=& 2 a_{i}^{(0)} \otimes a_{j}^{(0)} \otimes a_{k}^{(0)}
     \, - \, 2 a_{i}^{(0)} \otimes a_{j}^{(0)} \otimes a_{k}^{(0)}
     \, = \, 0 \, .
    \label{4-12}
\eeqar
Since $N_i$ and $a_{i}^{(\pm)}$ commute, the second and third terms vanish.
Further, from the definition (\ref{4-7}), we can easily find
\beq
    [A^{(\tau)}_{ij} , A^{(\tau)}_{ik} ] =
    [ N_{i} \otimes \d_{a_j} + N_{i} \otimes \bd_{a_j}, N_{i} \otimes \d_{a_k} + N_{i} \otimes \bd_{a_k} ]
    = 0 \, .
    \label{4-13}
\eeq
Similarly, we can easily find $[ \widetilde{A}_{jk} , \widetilde{A}_{ik} ] = 0$
by using $[ a_{k}^{(0)} , \d_{a_k} ] = [ a_{k}^{(0)} , \bd_{a_k} ] = 0$ and
$[A^{(\tau)}_{jk} , A^{(\tau)}_{ik} ]= 0$. The first equations are obvious since $a_{k}^{(0)}$
is independent of zero modes. The second relation can also be checked by
\beqar
    [A^{(\tau)}_{jk} , A^{(\tau)}_{ik} ] &=&
    [ N_{j} \otimes \d_{a_k} + N_{j} \otimes \bd_{a_k}, N_{i} \otimes \d_{a_k} + N_{i} \otimes \bd_{a_k} ]
    \nonumber \\
    &=&
    N_j \otimes N_i \otimes [ \d_{a_k} , \bd_{a_k} ] + N_j \otimes N_i \otimes [ \bd_{a_k} , \d_{a_k} ]
    \nonumber \\
    &=&  0 \, .
    \label{4-14}
\eeqar
Therefore the infinitesimal braid relations (\ref{4-8}), (\ref{4-9}) indeed hold
and we can properly define the holonomy operator for $\widetilde{A}$:
\beq
    \Theta_{R, \ga} (z, a) = \Tr_{R, \ga} \, \Path \exp \left[
    \sum_{r \ge 2} \oint_{\ga} \underbrace{ \widetilde{A} \wedge \widetilde{A}
     \wedge \cdots \wedge \widetilde{A}}_{r}
    \right] \, .
    \label{4-15}
\eeq

Utilizing the form of (\ref{4-7}), we can calculate
$ [ \widetilde{A}_{12} , \widetilde{A}_{23} ]$ as
\beqar
    \nonumber
    [ \widetilde{A}_{12} , \widetilde{A}_{23} ] &=&
    [ A_{12} , A_{23} ] + [ A^{(\tau)}_{12} , A^{(\tau)}_{23} ] \\
    \nonumber
    [ A^{(\tau)}_{12} , A^{(\tau)}_{23} ]  &=&
    N_1 \otimes \d_{a_2} \otimes \d_{a_3} + N_1 \otimes \d_{a_2} \otimes \bd_{a_3}
    \\
    && - N_1 \otimes \bd_{a_2} \otimes \d_{a_3} - N_1 \otimes \bd_{a_2} \otimes \bd_{a_3}
    \label{4-16}
\eeqar
where $[ A_{12} , A_{23} ]$ is given by (\ref{2-18}).
Thus, as in the case of (\ref{2-19}), the exponents of (\ref{4-15}) with path ordering $\Path$
can be expressed as
\beqar
    && \Path \sum_{r \ge 2}^{\infty} \oint_{\ga} \underbrace{\widetilde{A} \wedge
    \widetilde{A} \wedge \cdots \wedge \widetilde{A} }_{r}
    \nonumber \\
    &=& \sum_{r \ge 2}^{\infty} \oint_{\ga}  \widetilde{A}_{1 2} \widetilde{A}_{2 3}
    \cdots \widetilde{A}_{r 1}
    \, \om_{12} \wedge \om_{23} \wedge \cdots \wedge \om_{r 1}
    \nonumber \\
    &=& \sum_{r \ge 2}^{\infty} \frac{1}{4^{r+1}} \sum_{(h_1, h_2, \cdots , h_r)}
    (-1)^{h_1 + h_2 + \cdots + h_r} \,
    \nonumber \\
    && ~~~
    \times \,
    \bigg( a_{1}^{(h_1)} \otimes a_{2}^{(h_2)} \otimes \cdots \otimes a_{r}^{(h_r)}
     + \, \d_{a_1} \otimes \d_{a_2} \otimes \cdots \otimes \d_{a_r}
    \nonumber \\
    && ~~~~~~~~~~~~~~~~~~~~~~~~~~~~~~~~~~~~~~~~~~~
    + \, \mbox{(terms involving $\bd_i$'s)} \,
    \bigg)
    \nonumber \\
    && ~~~ \times
    \, \oint_{\ga} \om_{12} \wedge \om_{23} \wedge \cdots \wedge \om_{r1}
    \label{4-17}
\eeqar
where, in analogy to (\ref{2-20}), we define
$N_1 \otimes \d_{a_2} \otimes \d_{a_3} \otimes \cdots \otimes \d_{a_r}
\otimes \d_{a_1}$ as
\beq
    \hf [\d_{a_1} , N_1 ] \otimes \d_{a_2} \otimes \d_{a_3} \otimes \cdots \otimes \d_{a_r}
    =  \hf  \d_{a_1} \otimes \d_{a_2} \otimes \d_{a_3} \otimes \cdots \otimes \d_{a_r} \, .
    \label{4-18}
\eeq
The expression shows a natural extension of the original form (\ref{2-19}).
It gives a hybrid version of $a^{(\pm)}_{i}$'s and $\d_{a_i}$'s.
Thus we may consider the holonomy operator $\Theta_{R, \ga} (z,a)$ with (\ref{4-7}) as
a natural generalization of $\Theta_{R, \ga} (z)$ with zero mode variables.
In practical calculations, however,
we need to take account of the double periodicity of $z_i$'s and $a_i$'s.
We shall analyze this point in a moment but before doing so let us
consider another possibility for the choice of $A^{(\tau)}_{ij}$
which turns out to be more useful for the calculation of
the zero-mode part of $\Theta_{R, \ga} (z,a)$.

Notice that we make the terms that involve $\bd_i$'s in (\ref{4-17}) implicit
since, as discussed in the previous section, any physical observables of zero modes are
described by holomorphic functions in terms of $a_i$'s.
In other words, the conjugate derivatives $\bd_{a_i}$'s are auxiliary operators.
Motivated by this fact, we can assume another form of $A^{(\tau)}_{ij}$:
\beq
    A^{(\tau)}_{ij} =  1_i \otimes \d_{a_j} \, .
    \label{4-19}
\eeq
where $1_i$ is an identity that acts on the $i$-th Fock space $V_i$.
We may make the dimension of $1_i$ the same as that of the
$SL(2, {\bf C})$ representation $\rho$; thus this identity is equivalent to
those that appear in (\ref{2-7-1}).
By construction, $A_{ij}$ and $A_{ij}^{(\tau)}$ decouple to each other.
In fact, $A_{ij}^{(\tau)}$ can be interpreted as a $c$-number operator
since the constituents of $A_{ij}^{(\tau)}$ are now given by
$c$-numbers. This is obvious from the fact that the derivative
operators $\d_{a_i}$ can be replaced by $\frac{\pi \ba_i}{\im \tau}$.
In other words, there is no algebraic structure in $A_{ij}^{(\tau)}$
since it involves only the holomorphic derivatives.
This means that the commutator
$[ A^{(\tau)}_{ij} + A^{(\tau)}_{jk}, A^{(\tau)}_{ik} ]$ obviously vanishes.
Thus the infinitesimal braid relations for $\widetilde{A} = A + A^{(\tau)}$ automatically
reduce to those of the ``pure-gauge'' potential $A$ under the choice of (\ref{4-19}).
Therefore we can also define the holonomy operator in the form of
(\ref{4-15}), with $A^{(\tau)}_{ij}$ now redefined by (\ref{4-19}).

As long as we follow our definition of path ordering $\Path$
discussed in (\ref{2-15})-(\ref{2-21}), the holonomy operator $\Theta_{R, \ga} (z, a)$
automatically reduces to the pure-gauge operator $\Theta_{R, \ga} (z)$.
This can easily be seen from the fact that
the commutator $ [ \widetilde{A}_{12} , \widetilde{A}_{23} ]$,
reduces to $ [ A_{12} , A_{23} ]$ for the choice of (\ref{4-19}), being
in comparison with (\ref{4-16}).
Thus, in the present case, we need to
relax the meaning of the path ordering $\Path$
in order to extract zero-mode information out of $\Theta_{R, \ga} (z, a)$.
Remember that we have determined the path ordering $\Path$
such that it is suitable for the bialgebraic operators of photons
$A_{ij} = a_{i}^{(+)} \otimes a_{j}^{(0)} + a_{i}^{(-)} \otimes a_{j}^{(0)}$ where
$a_{i}^{(\pm)}$, $a_{i}^{(0)}$ obey the $SL(2, {\bf C})$ algebra.
The zero-mode bialgebraic operators $A_{ij}^{(\tau)} = 1_i \otimes \d_{a_j}$
are, however, essentially given by the differential operators $\d_{a_j}$
which are free from the $SL(2, {\bf C})$ algebra.
Thus it is natural to relax the meaning of the path ordering $\Path$
for the calculation of zero-mode information out of the holonomy operator
if we stick to the definition (\ref{4-19}).
(The change of algebraic properties in $\Theta_{R, \ga} (z, a)$ also suggests
that the zero-mode part of $\Theta_{R, \ga} (z, a)$ is
no longer a holonomy of conformal invariance but rather that of scale invariance.)
In the following, we shall redefine the path ordering $\Path$ such that we
can suitably extract the zero-mode part of $\Theta_{R, \ga} (z, a)$, which
we shall denote by $\Theta_{R, \ga}^{(\tau)} (z, a)$ from here on, with
$A_{ij}^{(\tau)}$ having the form of (\ref{4-19}).

%\vskip 0.5cm \noindent
%\underline{Calculation of $\Theta_{R, \ga} (z,a)$}
\vskip 0.5cm \noindent
\underline{Extraction of $\Theta^{(\tau)}_{R, \ga} (z,a)$}

Partly related to the redefinition of the path ordering,
there is a crucial condition for the calculation of $\Theta_{R, \ga} (z, a)$, {\it i.e.},
the double periodicity condition (\ref{3-1}) for $z_i$'s.
Remember that the complex coordinate on torus
is parametrized by $z = \xi_1 + \tau \xi_2$ where $0 \le \xi_r \le 1$
$(r= 1,2)$, with $\xi_r = 0$ and $\xi_r = 1$ being identified.
Since we set $\re \tau$ to zero, the complex coordinate
is then expressed as $z = \xi_1 + i \im \tau \xi_2$.
This means that $z$ can be parametrized by $\xi_1$
if we suitably scale $\im \tau$
or absorb $\xi_2$ into the definition of $\im \tau$.
Namely the imaginary part of $z$ can be controlled by $\im \tau$.
Thus, in our settings, $z$ is essentially parametrized by $0 \le \xi_1 \le 1$.

For an $n$-particle system, we have $n$ such parameters $z_i$ ($i = 1,2, \cdots n$).
Owing to the braid trace in $\Theta_{R, \ga} (z, a)$, the holonomy operator preserves
permutation invariance over the numbering index $i$. Thus, without losing generality,
we can impose the ordering condition
\beq
    0 \le z_1 \le z_2 \le \cdots \le z_n \le 1
    \label{4-20}
\eeq
where we consider $z_i$'s as real variables.
In the following, we assume this condition unless mentioned otherwise.
The number of elements for the permutation or the braid group $\B_n$ is $n-1$.
Thus we can fix one of $z_i$'s, say $z_1 = 0$.
Since the boundary values are identical, this means that we can further
fix $z_n$ to $z_n = 1$. Namely, we shall impose
\beq
    z_1 = 0 \, , ~~~ z_n = 1
    \label{4-21}
\eeq
on top of the ordering condition (\ref{4-20}).
This suggests that the path integral in $\Theta_{R, \ga} (z, a)$
can be carried out along the path $\ga$ defined by the line segment
\beq
    \ga = [ 0 , 1 ]
    \label{4-22}
\eeq
embedded in a complex plane.
This does not contradict with the fact that
the path $\ga$ is defined on the physical configuration space
$\C = {\bf C}^n / \S_n$ since $z_i$'s are now all on $\ga = [0, 1]$.
Because the boundaries of the line segment are identical,
the path $\ga$ also represents a closed path.

The condition (\ref{4-21}) is very stringent. For example, applying this to
the logarithmic one-form $\om_{ij} = d \log ( z_i - z_j )$, we find
\beqar
    \om_{1 j} &=& \frac{ d z_1 - d z_j }{ z_1 - z_j } = \frac{  d z_j }{ z_j}
    \, \equiv \, \om^{(0)}_{j} \, ,
    \label{4-23} \\
    \om_{n j} &=& \frac{ d z_n - d z_j }{ z_n - z_j } = \frac{ - d z_j }{1 - z_j}
    \, \equiv \, \om^{(1)}_{j} \, .
    \label{4-24}
\eeqar
{\it Our strategy is to express the zero-mode part of the holonomy operator, denoted by
$\Theta_{R, \ga}^{(\tau)} (z, a)$, in terms of
$\om^{(0)}_{j}$ and $\om^{(1)}_{j}$.}
As in (\ref{2-16}), the exponent of $\Theta_{R, \ga}(z,a)$ can be expanded as
\beq
    \sum_{r \ge 2}^{\infty} \oint_{\ga} \underbrace{ \widetilde{A} \wedge \widetilde{A}
    \wedge \cdots \wedge
    \widetilde{A}}_{r}
    = \sum_{r \ge 2}^{\infty} \oint_{\ga}  \sum_{ (i<j) }  \widetilde{A}_{i_1 j_1}
     \widetilde{A}_{i_2 j_2} \cdots  \widetilde{A}_{i_r j_r}
    \bigwedge_{k=1}^{r} \om_{i_k j_k}
    \label{4-25}
\eeq
where $\ga$ is specified by (\ref{4-22}) and
$(i < j)$ means that the set of indices $(i_1, j_1, \cdots , i_r , j_r)$
are ordered such that $1 \le i_1 < j_1 \le r$,$\cdots$,$1 \le i_r < j_r \le r$.
As mentioned earlier, action of the path ordering $\Path$ which has been
defined in (\ref{2-17}) does not give zero-mode information.
In order to obtain the zero-mode information,
we now introduce a new path ordering $\Path^{(\tau)}$ defined by
\beqar
    && \Path^{(\tau)} \sum_{r \ge 2} \oint_{\ga}
    \underbrace{ \widetilde{A} \wedge \widetilde{A} \wedge \cdots \wedge \widetilde{A}}_{r}
    \nonumber \\
    &=& \sum_{r \ge 2} \oint_{\ga}   \widetilde{A}_{1 2}
    \widetilde{A}_{1 3}   \cdots \widetilde{A}_{1 r} \widetilde{A}_{r 1}
    \, \om_{12} \wedge \om_{13} \wedge \cdots \wedge \om_{1r} \wedge \om_{r 1}
    \nonumber \\
    &=& \sum_{r \ge 2} \oint_{\ga}   \widetilde{A}_{1 2}
    \widetilde{A}_{1 3}   \cdots \widetilde{A}_{1 r} \widetilde{A}_{r 1}
    \, \om^{(0)}_{2} \wedge \om^{(0)}_{3} \wedge \cdots \wedge \om^{(0)}_{r} \wedge \om^{(1)}_{1}
    \, .
    \label{4-26}
\eeqar
The symbol $\Path^{(\tau)}$ means that, on top of the initial condition $(i < j)$,
the indices $(i_1, j_1, \cdots , i_r , j_r)$ are further constrained by
$i_1 = i_2 = \cdots  = i_{r-1} = 1$, $i_r = r$, and $2 \le j_1 < j_2 < \cdots < j_r \le r+1$
where $r+1$ is to be identified with $1$.
The ordering rule for $j$'s is the same as that of $\Path$; the difference lies in
the specific fixing of $i$'s. Notice that these extra conditions automatically
lead to the above expression (\ref{4-26}).

We now argue that the new path ordering $\Path^{(\tau)}$ in (\ref{4-26})
automatically leads to the zero-mode holonomy operator.
As mentioned below (\ref{4-19}), $A_{ij}^{(\tau)} = 1_i \otimes \d_{a_j}$
can be treated as a $c$-number operator which decouples with
$A_{ij} = a_{i}^{(+)} \otimes a_{j}^{(0)} + a_{i}^{(-)} \otimes a_{j}^{(0)}$.
Thus the factor of
$\widetilde{A}_{1 2} \widetilde{A}_{1 3}   \cdots \widetilde{A}_{1 r} \widetilde{A}_{r 1}$
in (\ref{4-26}) is split into the photon part
$A_{1 2} A_{1 3}   \cdots A_{1 r} A_{r 1}$ and the zero-mode part
$A^{(\tau)}_{1 2} A^{(\tau)}_{1 3}   \cdots A^{(\tau)}_{1 r} A^{(\tau)}_{r 1}$.
The vanishing of the photon-part can be
shown by the commutation relation
\beqar
    [ A_{1 2} , A_{1 3} ] &=&
    [ a_{1}^{(+)} \otimes a_{2}^{(0)} + a_{1}^{(-)} \otimes a_{2}^{(0)} , \,
    a_{1}^{(+)} \otimes a_{3}^{(0)} + a_{1}^{(-)} \otimes a_{3}^{(0)} ]
    \nonumber \\
    &=& 2 a_{1}^{(0)} \otimes a_{2}^{(0)} \otimes a_{3}^{(0)}
    - 2 a_{1}^{(0)} \otimes a_{2}^{(0)} \otimes a_{3}^{(0)} \, = \, 0 \, .
    \label{4-27}
\eeqar
On the other hand, the zero-mode part
$A^{(\tau)}_{1 2} A^{(\tau)}_{1 3} \cdots A^{(\tau)}_{1 r} A^{(\tau)}_{r 1}$
gives  a $c$-number operator acting on
the Hilbert space $V^{\otimes r} = V_1 \otimes V_2 \otimes \cdots \otimes V_r$ for zero modes.
Thus we can interpret this part as an coefficient of the loop integral in (\ref{4-26}).
Explicitly this factor can be calculated as
\beq
    A^{(\tau)}_{1 2} A^{(\tau)}_{1 3} \cdots A^{(\tau)}_{1 r} A^{(\tau)}_{r 1}
    =
    \d_{a_2} \otimes  \d_{a_3} \otimes \cdots \otimes \d_{a_r} \otimes \d_{a_1} \, .
    \label{4-27a}
\eeq
This statement is intuitively correct but is not mathematically rigorous
since, if we treat $A^{(\tau)}_{ij}$ as a bialgebraic operator which we do in
defining the (de)coupling between $A_{ij}$ and $A^{(\tau)}_{ij}$, the commutator
of $A^{(\tau)}_{ij}$'s obviously vanishes and, as in (\ref{4-27}), we can
argue that the zero-mode part of (\ref{4-26}) also becomes zero.
This problem originates from the fact that the physical operators in holonomy
formalism obey the $SL (2 , {\bf C})$ algebra.
For the zero-mode variables, however, there are no nontrivial algebraic
symmetries built-in as long as we use only the holomorphic part of the variables.
In this sense, the zero-mode holonomy operator does not represent a holonomy
operator of conformal invariance but rather scale invariance.
Thus it may be necessary to consider the commutator of $A^{(\tau)}_{ij}$'s
in constructing the full holonomy of $\widetilde{A} = A + A^{(\tau)}$.
But if one is interested in purely the holonomy of $A^{(\tau)}$,
it is sufficient to interpret $A^{(\tau)}_{ij}$ as a $c$-number operator.

Having said these rather intuitive justifications,
we now present a more satisfactory solution to the above problem, that is,
the problem can be fixed by
imposing an antisymmetric property on the derivative operators.
Since derivatives are commutative, it seems impossible to
impose such a condition. However, in the calculatory process
in (\ref{4-26}) it can be done by introducing
a coupling of the derivative operator $\d_{a_j}$ with a Grassmann
variable $\eta_j$. This variable is a ghost
variable that appears only in the middle of calculation and
will be integrated out at the end of calculation.
The operator $A^{(\tau)}_{ij}$ is then rewritten as
\beq
    A^{(\tau)}_{ij} = 1_i \otimes \int d \eta_j \, \eta_j \d_{a_j}
    = 1_i \otimes \d_{a_j}
    \label{4-27b}
\eeq
where $\eta_j$ is the Grassmann variable.
The final expression of $A^{(\tau)}_{ij}$ remains the same
but the above redefinition means that
we make it a rule to carry out the
Grassmann integral at the end of calculation.
Thus with an introduction of the ghosts, we can calculate
the quantity $A^{(\tau)}_{12}A^{(\tau)}_{13}$ as
\beqar
    A^{(\tau)}_{12}A^{(\tau)}_{13}
    &=& \int d \eta_3 d \eta_2 ~ 1_{1} \otimes \d_{a_2} \otimes \d_{a_3}
    \, \eta_2 \eta_3
    \nonumber \\
    &=& \frac{1}{2} \int d \eta_3 d \eta_2 ~ 1_{1} \otimes \d_{a_2} \otimes \d_{a_3}
    \, \left( \eta_2 \eta_3 - \eta_3 \eta_2 \right)
    \nonumber \\
    &=&
    1_{1} \otimes \d_{a_2} \otimes \d_{a_3} \, .
    \label{4-27c}
\eeqar
Similarly the factor the (\ref{4-27a}) can be expressed as
\beqar
    A^{(\tau)}_{1 2} A^{(\tau)}_{1 3} \cdots A^{(\tau)}_{1 r} A^{(\tau)}_{r 1}
    &=& \int [ d \eta ] ~ \d_{a_2} \otimes \d_{a_3} \otimes \cdots \otimes \d_{a_r}
    \otimes \d_{a_1} \,
    \eta_2 \eta_3 \cdots \eta_r \eta_1
    \nonumber \\
    &=&  \d_{a_2} \otimes \d_{a_3} \otimes \cdots \otimes \d_{a_r}
    \otimes \d_{a_1}
    \label{4-27d}
\eeqar
where $[ d \eta ] = d \eta_1 d \eta_r d \eta_{r-1} \cdots d \eta_2$.
This shows that the use of (\ref{4-27b}) guarantees
that the zero-mode part of (\ref{4-26}) gives non-vanishing contributions
even though $\d_{a_i}$ is a $c$-number operator.
Thus the expression (\ref{4-26}) can further be calculated as
\beqar
    && \Path^{(\tau)} \sum_{r \ge 2} \oint_{\ga}
    \underbrace{ \widetilde{A} \wedge \cdots \wedge \widetilde{A}}_{r}
    \nonumber \\
    &=& \sum_{r \ge 2}
     \d_{a_2} \otimes  \d_{a_3} \otimes \cdots \otimes \d_{a_r} \otimes \d_{a_1}  \, \oint_{\ga}
    \om^{(0)}_{2} \wedge \om^{(0)}_{3} \wedge \cdots \wedge \om^{(0)}_{r} \wedge \om^{(1)}_{1}
    \, . \nonumber \\
    \label{4-28}
\eeqar
As discussed above, this result can also be obtained by direct use of (\ref{4-27a}).
The introduction of the ghost variables is redundant in this sense
but they are necessary if one is interested in how the zero-mode contributions
incorporate into the abelian holonomy operators of conformal field theory.
In what follows we are interested in only the zero-mode part of the holonomy operator.
Thus we shall leave these calculatory issues aside in the rest of the present paper.

\vskip 0.5cm \noindent
\underline{Integral representation of Riemann's zeta function}

We now consider the integral part of the expression (\ref{4-28}).
This integral can be understood as an iterated integral of
the following one-forms:
\beqar
    \om^{(1)} &=& \frac{d t}{1 - t} \, \equiv \, f_1 (t) dt
    \label{r4-0}\\
    \om^{(0)} &=& \frac{d t}{ t } \, \equiv \, f_0 (t) dt
    \label{r4-1}
\eeqar
where $t \in \ga = [0,1]$.
It is known that the polylogarithm function $\Li_k (z)$ can be represented by
an iterated integral in general \cite{Kohno:2002bk}. For the case of $\Li_2 (1)$
this can be given by
\beq
    \int_\ga  \om^{(1)} \om^{(0)} = \int_{0}^{1} \frac{ \log ( 1- t)}{t} dt
    = \Li_2 (1) = \sum_{n=1}^{\infty} \frac{1}{n^2} = \frac{\pi^2}{6}
    \label{r4-2}
\eeq
where we use the definition of the iterated integral
\beq
    \int_\ga \om^{(1)} \om^{(0)} = \int_{0}^{1}
    \left( \int_{0}^{t_0} f_1 ( t_1 ) d t_1 \right)
    f_0 ( t_0 ) d t_0 \, .
    \label{r4-3}
\eeq
In general, higher-order iterated integrals in terms of one-forms
\beq
    \om_i = f_i (t) dt  ~~ (i = 1,2, \cdots , n)\, , ~~~ t \in \ga \, ,
    \label{r4-4}
\eeq
are defined by
\beq
    \int_\ga \om_1 \om_2 \cdots \om_n =
    \int_{\Delta_n} f_1 (t_{1}) f_2 (t_{2}) \cdots  f_n (t_{n})
    dt_1 dt_2 \cdots dt_n
    \label{r4-5}
\eeq
where $\Delta_n$ is given by
\beq
    \Delta_n = \left\{ (t_1 , t_2 , \cdots , t_n ) \in {\bf R}^n ~ | ~
    0 \le t_1 \le t_2 \le \cdots \le t_n \le 1 \right\} \, .
    \label{r4-6}
\eeq
Using the definition, we can generalize the expression (\ref{r4-2}) to obtain
\beq
    \Li_k (1) = \int_{0}^{1} \om^{(1)}
    \underbrace{\om^{(0)}  \om^{(0)} \cdots \om^{(0)}}_{k-1}
     = \sum_{n \ge 1} \frac{1}{n^k} = \zeta (k) \, .
    \label{r4-7}
\eeq
where $\zeta (k)$ is Riemann's zeta function.

Comparing our construction of
$\oint_\ga  \om^{(0)}_{2} \wedge \om^{(0)}_{3}
\wedge \cdots \wedge \om^{(0)}_{r} \wedge \om^{(1)}_{1}$
along with (\ref{4-20})-(\ref{4-24}) to the above definition
of iterated integrals, we find that
the loop integral can be calculated as
\beqar
    &&
    \oint_\ga  \om^{(0)}_{2} \wedge \om^{(0)}_{3}
    \wedge \cdots \wedge \om^{(0)}_{r} \wedge \om^{(1)}_{1}
    \nonumber \\
    &=& (-1)^{r-1} \oint_\ga   \om^{(1)}_{1} \wedge \om^{(0)}_{2} \wedge \om^{(0)}_{3}
    \wedge \cdots \wedge \om^{(0)}_{r}
    \nonumber \\
    &=& (-1)^{r} \int_\ga \om^{(1)} \underbrace{\om^{(0)}  \om^{(0)}  \cdots \om^{(0)}}_{r-1}
    \, = \, (-1)^{r} \zeta (r)
    \label{r4-8}
\eeqar
where we follow the notation of (\ref{r4-5}) in the last line.
The zero-mode holonomy operator is then expressed as
\beq
    \Theta_{R, \ga}^{(\tau)} (z, a) = \Tr_{R, \ga} \, \Path^{(\tau)} \exp \left[
    \sum_{r \ge 2} \oint_{\ga} \underbrace{ \widetilde{A} \wedge \widetilde{A}
     \wedge \cdots \wedge \widetilde{A}}_{r}
    \right]
    \label{4-29}
\eeq
where an explicit form of the exponent is given by
\beq
    \Path^{(\tau)} \sum_{r \ge 2} \oint_{\ga}
    \underbrace{ \widetilde{A} \wedge \cdots \wedge \widetilde{A}}_{r}
    = \sum_{r \ge 2} \,   \d_{a_1} \otimes
    \d_{a_2} \otimes  \d_{a_3} \otimes \cdots \otimes \d_{a_r} \, (-1)^{r} \zeta (r) \,
    \, .
    \label{r4-9}
\eeq
In the abelian case, the trace $\Tr_{R, \ga}$ is represented by a braid trace.
As in the expression (\ref{2-21}), this is
given by a sum over the permutation of the numbering indices.
We shall recapitulate
the final form $\Theta_{R, \ga}^{(\tau)} (z, a) \equiv \Theta_{R, \ga}^{(\tau)} (a)$
at the end of this section (see (\ref{4-37})).

\vskip 0.5cm \noindent
\underline{Generalization of linking numbers}

The vacuum expectation value (vev) of
the zero-mode holonomy operator $\Theta_{R, \ga}^{(\tau)} (z, a)$
or that of $\Theta_{R, \ga} (z, a)$ in general can be considered as a holomorphic function
of $(a_1 , a_2 , \cdots , a_n )$.
Thus, if we consider gauge transformations of these variables, {\it i.e.},
\beq
    a_i \, \rightarrow \, a_i + m_i + i \im \tau \, ,
    \label{4-30}
\eeq
we can certainly obtain analogs of linking numbers $m_i$ ($i= 1,2,\cdots, n$)
out of $\Theta_{R, \ga} (z ,a)$; in what follows we shall use
$\Theta_{R, \ga} (z ,a)$ for simplicity but this can always be
replaced by $\Theta_{R, \ga}^{(\tau)} (z ,a)$.
The corresponding holomorphic wavefunction, which is analogous to (\ref{3-17}), is then
written as
\beq
    \Xi [ \widetilde{A} ] \, = \, \exp \left( - \sum_{i = 1}^{n} \frac{K (a_i , \ba_i )}{2} \right)
    \bra \Theta_{R, \ga } (z ; a_1 , a_2 , \cdots , a_n ) \ket
    \label{4-31}
\eeq
where we make the $a_i$-dependence explicit.
The bracket $\bra \cdot \ket$ indicates that operators inbetween
are evaluated at the vacuum state of the system which is define on the
Hilbert space $V^{\otimes n}$ for zero modes.
The K\"{a}hler potential
$K ( a_i , \ba_i )$ is defined in (\ref{3-23}).
The polarization conditions for $\Xi [ \widetilde{A} ]$ are given by
\beq
    \left( \d_{a_i} + \frac{1}{2} \d_{a_i} K ( a_i , \ba_i ) \right)
    \, \Xi [ \widetilde{A} ] \, = \, 0
    \label{4-32}
\eeq
for arbitrary $i$'s.

The holonomy operator gives a monodromy representation of the KZ equation.
Thus it provides a general solution to the gauged KZ equation
\beq
    ( d - \widetilde{A} )  \Psi (z ; a_1 , a_2 , \cdots , a_n )  =  0
    \label{4-33}
\eeq
which is a differential equation analogous to (\ref{2-9}) with
$\widetilde{A}  =  A  +  A^{(\tau )}$.
This KZ equation is holomorphic in $a_i$'s, while the polarization condition (\ref{4-32})
is not. Thus, although these equations have similar structures,
there is no direct way to connect them.
Note that $z_i$ and $a_i$ are both coordinates on a torus, however,
the former is a coordinate of the Riemann surface on which two-dimensional conformal
field theory is defined, while the latter denotes a physical variable of zero modes on torus.
Thus physical meaning of the complex variables $z_i$, $a_i$ are distinct.
This also explains the qualitative difference between the equations (\ref{4-32}) and (\ref{4-33}).

Now, applying the result of (\ref{3-35}), we find the relation
\beq
    \bra \Theta_{R, \ga } (z , a ) \ket
    \, = \, (-1)^{m_1 + m_2 + \cdots m_n}
    \, \bra \Theta_{R, \ga } (z , a + m + i \im \tau ) \ket
    \label{4-34}
\eeq
where $m_i$ are integers and $\Theta_{R, \ga } (z , a + m + i \im \tau )$
is defined by
\beq
    \Theta_{R, \ga } (z ; a_1 + m_1 + i \im \tau , a_2 + m_2 + i \im \tau ,
    \cdots , a_n + m_n + i \im \tau ) \, .
    \label{4-35}
\eeq
As in the previous case, this relation can be checked by the gauge invariance
of the holomorphic wavefunction $\Xi [ \widetilde{A} ]= \Xi ( z, a )$ under
$a_i \rightarrow a_i + m_i + n_i \tau$ with $n_i = 1$, $\re \tau = 0$:
\beq
    e^{i \La_{m_i , 1} }\Xi ( z, a ) = \Xi ( z, a + m + i \im \tau )
    \label{4-36}
\eeq
where $\La_{m_i , 1 }$ is defined in (\ref{3-26}).
Notice that, as it will be obvious from the context,
$a_1$ and $a_2$ in (\ref{3-26}) do not denote zero-mode variables
for $i=1,2$ but correspond to the complex variables
defined in (\ref{3-10}) for arbitrary $i$'s.

\begin{figure} [htbp]
\begin{center}
\includegraphics[height=6.5cm]{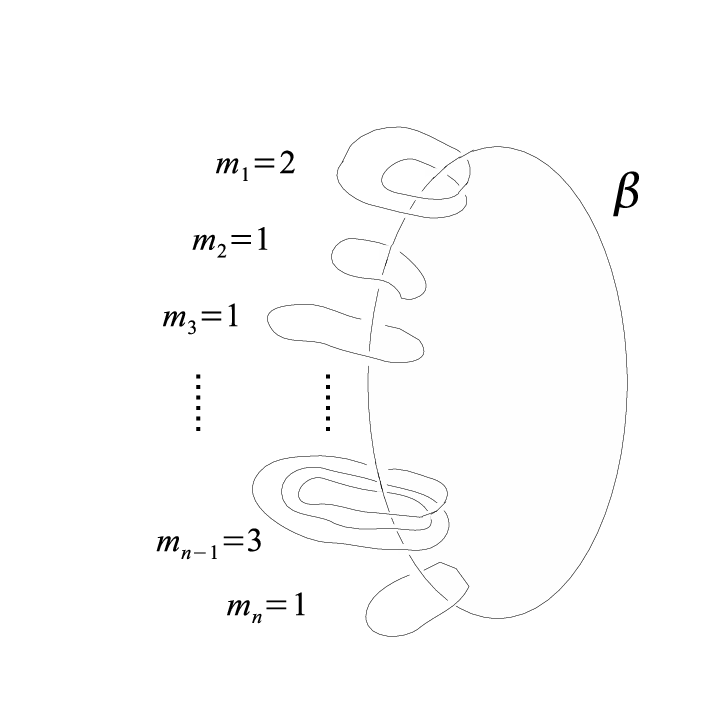}
\caption{How $\al$ cycles entangle with the common $\bt$ cycle.
The sum $m_1 + m_2 + \cdots + m_n$ in (\ref{4-34})
can be interpreted as a generalization of linking numbers
in the abelian holonomy formalism.}
\label{figNum02}
\end{center}
\end{figure}

The relation (\ref{4-34}) gives a straightforward generalization of (\ref{3-35}).
This means that we can consider the sum $m_1 + m_2 + \cdots + m_n$ in (\ref{4-34})
as a generalized linking number.
As discussed earlier, we can interpret $m_i$ as a linking number
of the $\al$ and $\bt$ cycles of the $i$-th torus on which
the zero-mode variable $a_i$ is defined.
The fact that we fix $n_i$ to 1 for arbitrary $i$'s suggests that we may
use a common $\bt$ cycle for each of $m_i$'s (see Figure \ref{figNum02}).

%%%%%%%%%%%%%%%%%%%%%%%%% figure %%%%%%%%%%%%%%%%%%%%%%%%%
%%%%%%%%%%%%%%%%%%%%%%%%% figure %%%%%%%%%%%%%%%%%%%%%%%%%

\vskip 0.5cm \noindent
\underline{Summary}

As a summary of this section, we recapitulate the resultant expression
for the zero-mode holonomy operator $\Theta^{(\tau)}_{R, \ga} (z,a)$ below.
\beqar
    \Theta_{R, \ga}^{(\tau)} (z, a) &=& \Tr_{R, \ga} \, \Path^{(\tau)} \exp \left[
    \sum_{r \ge 2} \oint_{\ga} \underbrace{ \widetilde{A} \wedge \widetilde{A}
     \wedge \cdots \wedge \widetilde{A}}_{r}
    \right] \nonumber \\
    &=& \Tr_{R, \ga} \exp \left[
    \sum_{r \ge 2} \,
    \d_{a_1} \otimes  \d_{a_2} \otimes \d_{a_3} \otimes  \cdots \otimes \d_{a_r} \,
    (-1)^{r} \zeta (r) \,
    \right]
    \nonumber \\
    &=& \exp \left[
    \sum_{r \ge 2}  \, \sum_{\si \in \S_{r-1}} \,  \d_{a_1} \otimes
    \d_{a_{\si_2}} \otimes \d_{a_{\si_3}} \otimes  \cdots \otimes \d_{a_{\si_r}} \,
    (-1)^{r} \zeta (r) \,
    \right]
    \nonumber \\
    & \equiv &  \Theta_{R, \ga}^{(\tau)} (a)
    \label{4-37}
\eeqar
where $\si$ denote permutations of the numbering elements $\{2,3, \cdots, r \}$,
{\it i.e.}, $\si=\left(%
\begin{array}{c}
  2 \, 3 \cdots r \\
  \si_2 \si_3 \cdots \si_r \\
\end{array}%
\right)$.
$\d_{a_i}$ ($i=1,2, \cdots, r$) are the derivative operators with respect to
the zero-mode variables $a_i$. These operators appear in the definition of
the comprehensive gauge field $\widetilde{A}$:
\beqar
    \widetilde{A}  &= & A  +  A^{(\tau )} \, ,
    \label{4-38}\\
    A^{(\tau)} &=& \sum_{1 \le i < j \le n} A^{(\tau)}_{ij} \om_{ij} \, ,
    \label{4-39} \\
    A^{(\tau)}_{ij} &=&  1_i \otimes \d_{a_j}
    \label{4-40}
\eeqar
where $A$ is defined in (\ref{2-23})-(\ref{2-25}).
Gauge transformations of $a_i$, {\it i.e.}, $a_i \rightarrow a_i + m_i + i \im \tau$,
in the vacuum expectation value of $\Theta_{R, \ga}^{(\tau)} (a)$ leads to the relation
\beq
    \bra \Theta_{R, \ga}^{(\tau)} ( a+ m + i \im \tau ) \ket \,
    = \, \prod_{i=1}^{n} (-1)^{m_i} \, \bra \Theta_{R, \ga}^{(\tau)} ( a) \ket
    \label{4-41}
\eeq
where the argument in the left-hand side means the same as that in (\ref{4-35}) and
$m_i \in {\bf Z}$ ($i = 1,2, \cdots , n)$ can be interpreted as a linking number
between the $\al_i$ cycle and the common $\bt$ cycle (see Figure \ref{figNum02}).
As mentioned in the introduction, the concept of linking numbers can be
related to the Legendre symbol of elementary number theory.
Motivated by this fact, in the next section we shall
consider $\Theta_{R, \ga}^{(\tau)} ( a)$
in a space of finite field ${\bf F}_p$ where $p$ is an odd prime number.

%%%%%%%%%%%%%%%%%%%%%%%%%%%%%%%%%%%%%%%%%%%%%%%%%
\section{Application to the elementary theory of numbers}

In this section we consider application of the zero-mode holonomy operator to
a space of finite field ${\bf F}_p$.
{\it As mentioned in the introduction, our strategy is to
use Morishita's result (\ref{1-1}) on the analogy between knots and primes.}
We first review this result and related materials in number theory
which are of direct relevance to later discussions.

\vskip 0.5cm \noindent
\underline{Linking numbers, Legendre symbols and Jacobi symbols}

For convenience and refreshment, we start writing down the result
(\ref{1-1}) again \cite{Morishita:2009gt}:
\beq
    ( -1 )^{lk (q, p)} = \left( \frac{q^*}{p} \right)
    \label{5-1}
\eeq
where $lk (q, p)$ is an analog of a linking number (mod 2) of
two distinct odd primes, $p, q$, and $\left( \frac{q^*}{p} \right)$
is the Legendre symbol defined in (\ref{1-2}).
The value of $q^{*}$ is given by $q^{*} = (-1)^{\frac{q-1}{2}} q$, {\it i.e.},
\beq
    q^*  = \left( \frac{-1}{q} \right)q =
    \left\{
    \begin{array}{l}
    q ~~~ \mbox{$q \equiv 1$ (mod 4)} \, ; \\
    - q ~~~ \mbox{$q \equiv 3$ (mod 4)} \, .  \\
    \end{array}
    \right.
    \label{5-2}
\eeq
The Legendre symbol satisfies the reciprocity law
\beq
    \left( \frac{q^*}{p} \right) = \left( \frac{p}{q} \right) \, .
    \label{5-3}
\eeq
In terms of $lk (p, q)$, this can also be written as
\beq
    lk (q, p) = lk (p , q^* ) = lk (p^* , q) \, .
    \label{5-4}
\eeq

%\underline{Gauss sum as a Fourier transform in mod $p$}
Now, denoting the Legendre symbel $\left( \frac{q^*}{p} \right)$ by $\la_{p} ( q^* )$,
we can express the Gauss sum as
\beq
    \widehat{\la}_p = \sum_{x = 1}^{p-1} \la_p (x) e^{i \frac{2\pi}{p}x}
    \label{5-5}
\eeq
where $x \in {\bf F}_{p}^{\times} = {\bf F}_p - \{ 0 \}$.
It is well-known that the Gauss sum becomes
\beq
    \widehat{\la}_p = \sqrt{p^*} =
    \left\{
    \begin{array}{l}
    \sqrt{p} ~~~ \mbox{$p \equiv 1$ (mod 4)} \, ; \\
    i \sqrt{p} ~~~ \mbox{$p \equiv 3$ (mod 4)} \, .  \\
    \end{array}
    \right.
    \label{5-6}
\eeq
For example, $\widehat{\la}_3$ and $\widehat{\la}_5$
can be calculated as
\beq
    \begin{array}{l}
    \widehat{\la}_3 = e^{i \frac{2 \pi}{3}} - e^{i \frac{4\pi}{3} }
    = i \sqrt{3} \, ,  \\
    \widehat{\la}_5
    = e^{i \frac{2 \pi}{5}} - e^{i \frac{4 \pi}{5}} - e^{i \frac{6 \pi}{5}} + e^{i \frac{8 \pi}{5}}
    = \sqrt{5} \, .
    \end{array}
    \label{5-7}
\eeq
This means that the Gauss sum takes a value of complex number
and that we can interpret the Legendre symbol as a map
$\la_p : \, {\bf F}_{p}^{\times}  \rightarrow  {\bf C}$.
The Gauss sum (\ref{5-5}) can then be interpreted as a
Fourier transform of $\la_p (x)$ in a space of mod $p$ \cite{Ono:1987bk}.
This interpretation is interesting but there is a caveat. Namely,
we may need to deal with a sum over all odd primes in order to
calculate the inverse Fourier transform, which is practically
impossible. However, calculation of the Gauss sum (\ref{5-5}), {\it per se},
does not involve such a sum and it will be useful to consider
(\ref{5-5}) as a Fourier transform particularly in applying
a field-theoretic approach to elementary number theory.
In this context, the ``phase space'' of interest is
given by $\left( \frac{2 \pi}{p} , x \right)$
with $x \in {\bf F}_{p}^{\times}$.
In a language of quantum field theory, this suggests that
the Legendre symbol $\la_p (x)$ can be interpreted as an operator
in an $x$-space representation, {\it i.e.}, an operator in a space of mod $p$.
In the same sense, the Gauss sum $\widehat{\la}_p$ can be regarded as
a conjugate operator, that is, we can interpret $\widehat{\la}_p$ as
an operator in a space of $\frac{2 \pi}{p}$ which is conjugate
to the space of ${\bf F}_{p}^{\times}$.
Thus it is presumably natural to interpret $\widehat{\la}_p$ as
an operator that is relevant to creation of primes.
We shall follow this idea later in application of the zero-mode holonomy operator
to number theory.

A generalization of the Legendre symbol does exist and it is called
the Jacobi symbol.
Let $N$ be an odd positive integer whose prime factorization is given by
\beq
    N = p_{1}^{e_1} p_{2}^{e_2} \cdots p_{s}^{e_s} \, .
    \label{5-8}
\eeq
Then the Jacobi symbol is defined by
\beq
     \left( \frac{y}{N} \right) =  \left( \frac{y}{p_1} \right)^{e_1}
     \left( \frac{y}{p_2} \right)^{e_2} \cdots  \left( \frac{y}{p_s} \right)^{e_s}
    \label{5-9}
\eeq
where $y$ denotes an arbitrary integer.
In terms of $\la_p (y)$, this can also be expressed as
\beq
    \left( \frac{y}{N} \right) = { \la_{p_1} (y) }^{e_1} { \la_{p_2} (y) }^{e_2} \cdots
    { \la_{p_s} (y) }^{e_s} := \eta_{\mbox{{\tiny $N$}}} (y) \, .
    \label{5-10}
\eeq
It is known that the Jacobi symbol also satisfies the supplementary and reciprocity laws
which are essentially the same as those of the Legendre symbol (see, {\it e.g.}, \cite{Ono:1987bk}).
In number theory, the Jacobi symbols are particularly useful in calculation
of the Legendre symbol $\left( \frac{x}{p} \right)$ where $p$ is a large prime number.

\vskip 0.5cm \noindent
\underline{Correspondence between knots and primes}
%\underline{Generating function for prime factorization of integers}

We now consider the zero-mode holonomy operator
$\Theta_{R, \ga}^{(\tau)} (a)$ in a space of ${\bf F}_{p}^{\times}$, utilizing
the above results.
As discussed above, quantum theoretically
the Legendre symbol can be interpreted as an operator in this space.
On the other hand, the operative construction of
$\Theta_{R, \ga}^{(\tau)} (a)$ is summarized in (\ref{4-37})
and its vacuum expectation values satisfy the relation (\ref{4-41}).
Thus a natural way to carry out our analysis is
to interpret the factor of $(-1)^{m_i}$ in (\ref{4-41}) as a Legendre symbol.

As discussed in (\ref{3-33}), a linking number of
the $\al_i$ and $\bt_i$ cycles can be defined by
\beq
    m_i n_i = lk ( \al_{i}^{m_i} , \bt_{i}^{n_i} )
    \label{5-11}
\eeq
where $m_i$ and $n_i$ are the winding numbers of $\al_i$ and $\bt_i$
cycles, respectively. We make the latter being fixed to $n_i = 1$.
Furthermore, as shown in Figure \ref{figNum02}, we have
made $\bt_i$ identical for arbitrary $i$'s.
Thus in our settings the winding number $m_i$ can be
written as
\beq
    m_i = lk ( \al_{i}^{m_i} , \bt ) \, .
    \label{5-12}
\eeq
In this section, we further impose
\beq
    m_i = 1
    \label{5-13}
\eeq
so that we can single out the linking number of
the $\al_i$ cycle and the common $\bt$ cycle
in the following form:
\beq
    \left. ( - 1 )^{m_i} \right|_{m_i = 1}
    = (-1 )^{ lk (\al_i ,  \bt )} \, .
    \label{5-14}
\eeq
According to Morishita's analogies between knots and primes,
the $\al_i$ and $\bt$ cycles correspond to odd prime numbers.
Thus we may express the above quantity as
\beqar
    && (-1)^{ lk (\al_{i} , \bt )}
    = \la_{\bt} ( \al_{i}^{*} ) = \la_{\al_i} ( \bt )
    \nonumber \\
    &=& ( -1 )^{lk ( \bt , \al_{i}^{*} )}
    = \la_{ \al_{i}^{*} } ( \bt^* ) = \la_{\bt } ( \al_{i}^{*} )
    \label{5-15} \\
    &=& ( -1 )^{lk ( \bt^* , \al_i )}
    = \la_{ \al_i } ( \bt ) = \la_{\bt^* } ( \al_i )
    \nonumber
\eeqar
where we use (\ref{5-4}). These relations hold when
$\al_i$ and $\bt$ are all odd prime numbers.
Notice, however, that we are going to consider the Legendre symbol
as a {\it function} of $x \in {\bf F}^{\times}_{p}$.
As we mentioned earlier, we shall consider the Legendre symbol as
an {\it operator} of $x \in {\bf F}^{\times}_{p}$ eventually (see (\ref{5-27}));
here we shall make a classical analysis for the moment.
Since the elements of ${\bf F}_{p}^{\times}$ include non-prime numbers,
the relations (\ref{5-15}) are not quite applicable to our settings
but they do suggest that we have several (essentially two, as discussed below)
interpretations to the Legendre symbols
in terms of the correspondence between knots and primes.
For example, using the relation $(-1)^{ lk (\al_{i} , \bt )}
 = \la_{\al_i} ( \bt )$ in (\ref{5-15}), one can
interpret $\al_i$ as an odd prime number and $\bt$ as an integer
in ${\bf F}^{\times}_{\al_i}$.
Another example is to
use the relation $( -1 )^{lk ( \bt^* , \al_i )}
=  \la_{\bt^* } ( \al_i )$ and to
interpret $\al_i$ as an integer
in ${\bf F}^{\times}_{\bt^*}$ and $\bt$ as an odd prime number.
By a suitable choice of $\al_{i}$ and $\bt$, we can in fact
classify the Legendre symbols of (\ref{5-15}) into two types,
{\it e.g.}, $\la_{\al_i} ( \bt )$ and $\la_{\bt^* } ( \al_i )$,
depending on the ordering of $\al_i$ (or $\al_{i}^{*}$) and $\bt$ (or $\bt^*$).
Note here that for $p \equiv 3$ (mod 4), we have $p^* = - p$. Thus we may not
properly define the Legendre symbol $\la_{p^*} (x)$ $(x \in {\bf F}_{p^*}^{\times})$.
However, as far as the calculation of the Gauss sum (\ref{5-6}) is concerned,
we can properly define it with $p^*$. We shall discuss this point later
in (\ref{5-36}).

%\vskip 0.5cm \noindent
%\underline{Jacobi symbols and zero-mode holonomy operators}
Using the interpretation of $(-1)^{ lk (\al_{i} , \bt )}
 = \la_{\al_i} ( \bt )$, we now illustrate a simple connection
between the Jacobi symbol and the vacuum expectation values of
the zero-mode holonomy operator.
For the choice of $m_i = 1$, the factor of $( -1 )^{m_i}$ becomes
$\la_{\al_i} (\bt )$. Thus the relation (\ref{4-41}) can be written as
\beqar
    \bra \Theta_{R, \ga}^{(\tau)} ( a+ 1 + i \im \tau ) \ket
    & = &
    \prod_{i}^{n} \la_{ \al_i} (\bt ) \, \bra \Theta_{R, \ga}^{(\tau)} ( a) \ket
    \nonumber \\
    &=&
    \left( \frac{\bt}{\al_{1} \al_{2} \cdots \al_{n} } \right)
    \, \bra \Theta_{R, \ga}^{(\tau)} ( a) \ket
    \nonumber \\
    &=&
    \eta_{\mbox{{\tiny $N_1$}}} (\bt)
    \, \bra \Theta_{R, \ga}^{(\tau)} ( a) \ket
    \label{5-16}
\eeqar
where we use the notation (\ref{4-35}) and
$\eta_{\mbox{{\tiny $N_1$}}} (\bt)$ is the Jacobi symbol, with
$\al_i$ ($i = 1, 2, \cdots , n$) and $\bt$ denoting odd primes and
an arbitrary integer, respectively. $N_1$ is an odd integer defined by
\beq
    N_1 = \al_1 \al_2 \cdots \al_n \, .
    \label{5-17}
\eeq

The general form of the Jacobi symbol, in the form of (\ref{5-10}),
can also be obtained by relaxing the choice of winding numbers $n_i \in {\bf Z}$.
Let elements of $\{ p_1 , p_2 , \cdots , p_s \} $ be
\beq
    \{ p_1 , p_2 , \cdots , p_s \} \in \{ \al_1 , \al_2 , \cdots , \al_n \}
    ~~( s \le n ) \, ,
    \label{5-18}
\eeq
say, $\{ p_1 , p_2 , \cdots , p_s \}
= \{ \al^{\prime}_{1} , \al^{\prime}_{2} , \cdots , \al^{\prime}_{s} \}$,
and let the corresponding winding numbers of the $\bt$ cycle be
\beq
    \{ e_{1} , e_{2} , \cdots , e_{s} \} \in \{ n_1 , n_2 , \cdots , n_n \} \, ,
    \label{5-19}
\eeq
say, $\{ e_1 , e_2 , \cdots , e_s \}
= \{ n^{\prime}_{1} , n^{\prime}_{2} , \cdots , n^{\prime}_{s} \}$ $( s \le n )$,
with the rest of $n_i$'s being zero.
In this case, non-vanishing ``phase factors'' in (\ref{4-41}) are given
by
\beq
    ( - 1 )^{lk ( \al^{\prime}_{j} , \, \bt^{n^{\prime}_{j}} )}=
    \la_{\al^{\prime}_{j}} ( \bt^{n^{\prime}_{j}} )
    = \left( \frac{\bt^{n^{\prime}_{j}}}{\al^{\prime}_{j}} \right)
    = \left( \frac{\bt}{\al^{\prime}_{j}} \right)^{n^{\prime}_{j}}
    \label{5-20}
\eeq
$( j= 1,2, \cdots, s \le n )$ where we use
the multiplicative property of the Legendre symbols
\beq
    \left( \frac{x_1 x_2}{p} \right)
    = \left( \frac{x_1}{p } \right) \left( \frac{x_2}{p} \right) \, .
    \label{5-21}
\eeq
Then the general form of the Jacobi symbol arises from
the vacuum expectation value of the zero-mode holonomy operator as follows:
\beqar
    \bra \Theta_{R, \ga}^{(\tau)} ( a+ 1 + \sqrt{-1} n^\prime \im \tau ) \ket
    & = &
     \la_{\al^{\prime}_{1}} (\bt )^{n^{\prime}_{1}}
     \la_{\al^{\prime}_{2}} (\bt )^{n^{\prime}_{2}} \cdots
     \la_{\al^{\prime}_{s}} (\bt )^{n^{\prime}_{s}}
    \, \bra \Theta_{R, \ga}^{(\tau)} ( a ) \ket
    \nonumber \\
    &=&
    \eta_{\mbox{{\tiny $N$}}} (\bt)
    \, \bra \Theta_{R, \ga}^{(\tau)} ( a) \ket
    \label{5-22}
\eeqar
where we use (\ref{5-10}).
The argument of $\Theta_{R, \ga}^{(\tau)} ( a+ 1 + \sqrt{-1} n^\prime \im \tau )$
means that $m_i = 1$ and $n^{\prime}_{j} = 1$, with the rest of $n_i$'s being zero.
$N$ is an arbitrary odd integer represented by the prime factorization
\beq
    N = {\al^{\prime}_{1}}^{n^{\prime}_{1}}
    {\al^{\prime}_{2}}^{n^{\prime}_{2}} \cdots {\al^{\prime}_{s}}^{n^{\prime}_{s}}
    = p_{1}^{e_1} p_{2}^{e_2} \cdots  p_{s}^{e_s} \, .
    \label{5-23}
\eeq
This shows that, even at classical level,
we can extract information on prime factorization of integers
(or irreducible representation of integers)
in terms of the zero-mode holonomy operator applied to the knot-prime
correspondence.
Thus the result (\ref{5-22}) illustrates
a classical realization of the concept in (\ref{1-5}).
In what follows, we shall consider quantum aspects of this concept.
It will turn out that the quantum aspects are something more profound than
the classical ones discussed above.

\vskip 0.5cm \noindent
\underline{Quantum realization of linking numbers}

From the relation (\ref{3-31}), we find that the operator
$\d_{a_i}$ in $\Theta_{R, \ga}^{(\tau )} (a)$ can be
replace by $\frac{\pi}{\im \tau} \ba_i$ in the $\ba_i$-representations.
Under the transformation of $a_i \rightarrow a_i + m_i + i \im \tau$,
this factor changes as
\beq
    \frac{\pi}{\im \tau} \ba_i \, \longrightarrow \,
    \frac{\pi}{\im \tau} \left( \ba_i + m_i - i \im \tau \right)
    = \frac{\pi}{\im \tau} \ba_i - i \pi \left( 1 + i \frac{m_i}{\im \tau} \right) \, .
    \label{5-24}
\eeq
Substituting this relation into (\ref{4-41}), we find
\beq
    ( - 1 )^{m_i} \leftrightarrow
    \left\langle e^{-i \pi \left( 1 + i \frac{m_i}{\im \tau} \right)} \right\rangle \, .
    \label{5-25}
\eeq
By the choice of $m_i = 1$ as in (\ref{5-14}), this relation reduces to
\beq
    \left. ( - 1 )^{m_i} \right|_{m_i = 1} = ( - 1 )^{lk ( \al_i , \bt)}
    \leftrightarrow \left\langle (-1)^{ 1 +  \frac{i}{ \im \tau} } \right\rangle \, .
    \label{5-26}
\eeq

Since $\im \tau$ is simply a real parameter, the vacuum expectation value
in (\ref{5-26}) does not make sense unless we consider the factor $(-1)$ as some operator.
Naturally, this operator can be identified with the Legendre symbol.
This will be obvious from the correspondence between knots and primes
discussed in (\ref{5-14}) and (\ref{5-15}).
As mentioned below (\ref{5-15}), there are essentially two interpretations
to the Legendre symbols in connection with the linking number.
Here we shall consider the following two cases:
\beq
    ( -1 ) = \left\{
    \begin{array} {c}
    (-1)^{lk ( \al_i , \bt )} = \la_{ \al_i } ( \bt )  \, ; \\
    (-1)^{lk ( \bt^* , \al_i )} = \la_{\bt^*} ( \al_i )   \, .
    \end{array}
    \right.
    \label{5-27}
\eeq
The first interpretation is to consider
$\al_i$ as an odd prime number and $\bt$ as an integer
($x \in {\bf F}^{\times}_{\al_i}$), while the second one is to
consider $\bt$ as an odd prime and
$\al_i$ as an integer ($x \in {\bf F}^{\times}_{\bt^*}$).
The first case involves multiple prime numbers for $i= 1,2, \cdots, n$
and this will lead to give a quantum or operative
version of the relation in (\ref{5-16}).
On the other hand, the second case involves a single prime number and
this case will be more convenient to deal with ``dynamics'' of primes
in the zero-mode holonomy formalism, particularly in relation with Riemann's zeta function.
In the following, we shall study details of these considerations.

\vskip 0.5cm \noindent
\underline{Analogs of scattering amplitudes for prime numbers}
%\underline{Analogs of scattering amplitudes for integers}

We begin with the first case of (\ref{5-27}).
From (\ref{4-41}) and (\ref{5-26}), we can
express the ``gauged'' abelian holonomy operator
$\Theta_{R, \ga}^{(\tau)} ( a+ 1 + i \im \tau )$ as
\beqar
    && \Theta_{R, \ga}^{(\tau)} ( a + 1 + i \im \tau ; \al_i , \bt )
    \nonumber \\
    &=& \Tr_{R, \ga } \, \exp \Bigg[
    \sum_{r \ge 2} \, \left( \la_{\al_1} (\bt ) \la_{\al_2} (\bt)
    \cdots \la_{\al_r} ( \bt ) \right)^{1+ \frac{i}{\im \tau}}
    \nonumber \\
    && ~~~~~~~~~~~~~~~~~~~~~~~~~~~~~~~~~~~~~~~~~~~~~~
    \,    \d_{a_1} \otimes \d_{a_2} \otimes \cdots \otimes \d_{a_r}
    \, (-1)^{r}\zeta (r)
    \Bigg]
    \nonumber \\
    &=&
    \exp \Bigg[
    \sum_{r \ge 2} \, \sum_{ \si \in \S_{r-1}} \,
    \eta_{\mbox{{\tiny $N_1$}}} (\bt)^{1+ \frac{i}{\im \tau}}\,  \d_{a_1} \otimes
    \d_{a_{\si_2}} \otimes \d_{a_{\si_3}} \otimes  \cdots \otimes \d_{a_{\si_r}} \,
    (-1)^{r} \zeta (r) \,
    \Bigg] \, .
    \nonumber \\
    \label{5-28}
\eeqar
Here $N_1$ is given by $N_1 = \al_1 \al_2 \cdots \al_r$,
with $\al_i$ ($i= 1,2,\cdots , r$) denoting distinct odd prime numbers.
An analog of a scattering amplitude of these primes, {\it i.e.}, the
irreducible constituents of the odd integer $N_1$, is then written as
\beqar
    && \left. \A ( N_1 , \bt ) \right|_{N_1 = \al_1 \al_2 \cdots \al_n }
    \nonumber \\
    &\equiv&
    \left.
    \frac{1}{(n-1)!}
    \frac{\del}{\del \d_{a_1}} \otimes \frac{\del}{\del \d_{a_2}}
    \otimes \cdots \otimes
    \frac{\del}{\del \d_{a_n}}
    \Theta_{R, \ga}^{(\tau)} ( a + 1 + i \im \tau ; \al_i , \bt )
    \right|_{\d_a = 0}
    \nonumber \\
    &=&
    \eta_{\mbox{{\tiny $N_1$}}} (\bt)^{1+ \frac{i}{\im \tau}} \,
    (-1)^{n}\zeta (n)
    \label{5-29}
\eeqar
where $\d_a = 0$ in the second line means that the remaining
$\d_{a_i}$ operators (or source functions to be precise)
all vanish upon the completion of the functional derivatives.
Notice that this expression is in the same form as the
scattering amplitudes of photons in the holonomy formalism
except that we do not use physical operators $a_{i}^{\pm}$
but the zero-mode operators $\d_{a_i}$ here.
(For the description of holonomy operators as S-matrix functionals
of scattering amplitudes for gluons in general, see \cite{Abe:2009kn}.)
Thus we can naturally interpret
$\eta_{\mbox{{\tiny $N_1$}}} (\bt)^{1+ \frac{i}{\im \tau}} \zeta (n)$
as an analog of scattering amplitude for $n$ prime numbers, with
their values ($\al_1 , \al_2 , \cdots , \al_n$) not being specified.
In this sense, the zero-mode holonomy operator gives
a generating function for the ``scattering amplitudes'' of prime numbers.
As mentioned in the introduction, obtaining the expression
(\ref{5-29}) is a main goal of the present paper.

Any physical observable is given by the square of a probability amplitude.
The ``scattering probability'' corresponding to $\A ( N_1 , \bt )$ is then
expressed, up to normalization, as
\beq
    | \A ( N_1 , \bt ) |^2   \, = \, \zeta (n)^2
    \label{5-30}
\eeq
where we use the fact that the value of the Jacobi symbol
$\eta_{\mbox{{\tiny $N_1$}}} (\bt)$ is nothing but $\pm 1$;
note that each of the Legendre symbols $\la_{\al_i} (\bt)$ takes
a value of $\pm 1$.

The expressions (\ref{5-29}) and (\ref{5-30}) are independent of the choices of
$n$ prime numbers ($\al_1 , \al_2 , \cdots , \al_n$).
In fact, we can calculate these quantities even with
$\al_i = 1$ ($i = 1,2, \cdots, n$):
\beqar
    && \left. \A ( N_0 , \bt )
    \right|_{N_0 = \underbrace{1 \times 1 \times \cdots \times 1}_{n} }
    \nonumber \\
    &\equiv&
    \left.
    \frac{1}{(n-1)!}
    \frac{\del}{\del \d_{a_1}} \otimes \frac{\del}{\del \d_{a_2}}
    \otimes \cdots \otimes
    \frac{\del}{\del \d_{a_n}}
    \Theta_{R, \ga}^{(\tau)} ( a + 1 + i \im \tau ; \al_i , \bt )
    \right|_{\d_a = 0}
    \nonumber \\
    &=&
    \eta_{\mbox{{\tiny $N_0$}}} (\bt)^{1+ \frac{i}{\im \tau}} \,
    (-1)^{n} \zeta (n)
    ~=~ (-1)^{n} \zeta (n)
    \label{5-31}
\eeqar
where we use the conventional definition of the Jacobi symbol
\beq
    \eta_{\mbox{{\tiny 1}}} ( \bt ) \, = \, \left( \frac{\bt}{1} \right) \, = \,  1 \, .
    \label{5-32}
\eeq
Notice that the result (\ref{5-31}) can also be obtained
by use of $\Theta_{R, \ga}^{(\tau)} (a)$ in (\ref{4-37}), the ``pure-gauge'' holonomy operator.
This confirms that we can properly understand
$\Theta_{R, \ga}^{(\tau)} ( a + 1 + i \im \tau ; \al_i , \bt)$
as a gauged operator of $\Theta_{R, \ga}^{(\tau)} ( a ; \al_i , \bt )$.

As in the classical case, a generalization of (\ref{5-29}) can be
carried out by relaxing the condition for the winding numbers $n_i$.
This leads to the expression
\beqar
    && \!\!\! \left. \A ( N , \bt )
    \right|_{N = p_{1}^{e_1} p_{2}^{e_2} \cdots p_{s}^{e_s}}
    \nonumber \\
    &\equiv& \!\!\!
    \left.
    \frac{1}{(n-1)!}
    \frac{\del}{\del \d_{a_1}} \otimes \frac{\del}{\del \d_{a_2}}
    \otimes \cdots \otimes
    \frac{\del}{\del \d_{a_n}}
    \Theta_{R, \ga}^{(\tau)} ( a + 1+ \sqrt{-1} n^{\prime} \im \tau ; \al_i , \bt )
    \right|_{\d_a = 0}
    \nonumber \\
    &=& \!\!\!
    \eta_{\mbox{{\tiny $N$}}} (\bt)^{1+ \frac{i}{\im \tau}} \,
    (-1)^{n}\zeta (n)
    \label{5-33}
\eeqar
where we follow the same notations in (\ref{5-18})-(\ref{5-23}).
The resultant ``scattering amplitude'' is essentially same as (\ref{5-29})
except that now the prime factorization of the integer $N$ is given by
(\ref{5-23}) rather than that of $N_1$ in (\ref{5-17}).
The squares of these amplitudes are all identical to $\zeta (n)^2$, {\it i.e.},
\beq
    | \A ( N_0 , \bt ) |^2
    \, = \,
    | \A ( N_1 , \bt ) |^2
    \, = \,
    | \A ( N , \bt )  |^2
    \, = \,  \zeta (n)^2 \, .
    \label{5-34}
\eeq
This means that the normalized scattering probability is always one.
Physically this is obvious because
once we choose $N$ (and $\bt$) to calculate the amplitudes,
the irreducible representation of $N$, {\it i.e.}, a set of the values
of $\al_i$, $n_i$ (and $m_i =1$), is determined by hand
so that there is no notion of probability in the existence of $N$.
In other words, once an odd integer which factorizes into odd primes is given,
there arises no notion of creation for the integer; it is already there.
Thus it is probably not appropriate to calculate quantities such
as scattering probabilities of primes involving a particular prime
factorization in a framework of quantum field theory.

One may consider that the above argument can be improved
by integrating out the $\bt$-dependence.
This means the use of the Gauss sum for each of $\la_{\al_i} (\bt )$
in (\ref{5-28}).
In calculating the Gauss sum, we need to sum over
$\bt \in {\bf F}_{\al_i}^{\times}$ for each of the primes $\al_i$.
However, since there are $n$ distinct such primes, it is impossible to
execute a set of summations with a single parameter $\bt$; we need
multiple $\bt$'s corresponding to distinct $\al_i$'s.
Thus the $\bt$-dependence can not be integrated out in
the present interpretation of $\al_i$ and $\bt$.
This becomes possible if we choose the other interpretation of
$\al_i$ and $\bt$, namely, the second interpretation of (\ref{5-27}).
We shall consider this case in the following.

\vskip 0.5cm \noindent
\underline{The Gauss sum as a prime-creation operator}

Applying the second interpretation of (\ref{5-27}) to the expression
(\ref{5-28}), we can similarly write down the gauged abelian
holonomy operator $\Theta_{R, \ga}^{(\tau)} ( a+ 1 + i \im \tau )$ as
\beqar
    && \Theta_{R, \ga}^{(\tau)} ( a + 1 + i \im \tau ; \bt , \al_i )
    \nonumber \\
    &=& \Tr_{R, \ga } \, \exp \Bigg[
    \sum_{r \ge 2} \, \left( \la_{\bt^*} (\al_1 ) \la_{\bt^*} (\al_2 )
    \cdots \la_{\bt^*} (\al_r )  \right)^{1+ \frac{i}{\im \tau}}
    \nonumber\\
    && ~~~~~~~~~~~~~~~~~~~~~~~~~~~~~~~~~~~~
    \,    \d_{a_1} \otimes \d_{a_2} \otimes \cdots \otimes \d_{a_r}
    \,(-1)^{r} \zeta (r)
    \Bigg]
    \nonumber \\
    &=&
    \exp \Bigg[
    \sum_{r \ge 2} \, \sum_{ \si \in \S_{r-1}} \,
    \left( \la_{\bt^*} (\al_1 ) \la_{\bt^*} (\al_2 )
    \cdots \la_{\bt^*} (\al_r )  \right)^{1+ \frac{i}{\im \tau}}
    \nonumber\\
    && ~~~~~~~~~~~~~~~~~~~~~~
    \,  \d_{a_1} \otimes \d_{a_{\si_2}} \otimes \d_{a_{\si_3}}
    \otimes  \cdots \otimes \d_{a_{\si_r}} \, (-1)^{r} \zeta (r) \,
    \Bigg]
    \label{5-35}
\eeqar
where $\bt$ is now an odd prime and $\al_i $ ($i= 1,2, \cdots , r$) is an integer
defined in a space of ${\bf F}_{\bt^*}^{\times}$.
Thus, in this case, we can suitably consider a Fourier
transform of $\la_{\bt^*} ( \al_i )$ for each $i$.
This means that we can replace each of $\la_{\bt^*} ( \al_i )$ by
\beq
    \widehat{\la}_{\bt^*} \, = \, \sum_{\al_i \in {\bf F}_{\bt^*}^{\times}}
    \la_{\bt^*} ( \al_i ) e^{i \frac{2 \pi}{\bt^*} \al_i} \, = \, \sqrt{\bt}
    \label{5-36}
\eeq
where we naively use the relations in (\ref{5-5}) and (\ref{5-6}), that is,
although the Legendre symbol $\la_{\bt^*} ( \al_i )$
($\al_i \in {\bf F}_{\bt^*}^{\times}$) is not appropriately defined
for $\bt \equiv 3$ (mod 4), the result $\widehat{\la}_{\bt^*} = \sqrt{\bt}$
holds for any odd prime $\bt$ by a direct application of (\ref{5-6}).
In fact, a naive calculation of $\widehat{\la}_{3^*}$ suggests that
$\la_{3^*} ( x )$ ($\al_i \in {\bf F}_{3^*}^{\times}$)
should be defined by $\sqrt{-1}$.
Thus, practically speaking, it is not possible to define $\la_{\bt^*} ( \al_i )$
in compatible with the definition of the Legendre symbol (\ref{1-2}).
However, as far as the calculation of the Gauss sum is concerned,
we can bypass these issues and directly use the result in (\ref{5-36}).

Using (\ref{5-36}), we can properly integrate out
the $\al_i$-dependence in (\ref{5-35}) to obtain
\beqar
    && \Theta_{R, \ga}^{(\tau)} ( a + 1 + i \im \tau ; \bt )
    \nonumber \\
    &=&
    \!\!\!
    \Tr_{R, \ga } \, \exp \left[
    \sum_{r \ge 2} \, \sqrt{\bt}^{ \left( 1+ \frac{i}{\im \tau} \right) r}
    \, \d_{a_1} \otimes \d_{a_2} \otimes \cdots \otimes \d_{a_r}
    \, (-1)^{r} \zeta (r)
    \right]
    \label{5-37}
\eeqar
where we express the braid trace by $\Tr_{R, \ga}$ for simplicity.
As mentioned earlier, below (\ref{5-7}) and also in the introduction,
the Gauss sum (\ref{5-36}) can be interpreted as an operator
in a space of $\frac{2 \pi}{\bt^*}$.
It is a conjugate operator of the Legendre symbol $\la_{\bt^*} ( \al_i )$
which can, on the other hand, be considered as an operator
in a space of $\al_i \in {\bf F}_{\bt^*}^{\times}$.
Notice that, in a language of quantum field theory, the Fourier transform
(\ref{5-36}) shows that the relevant phase space for these operators is
given by $\left( \frac{2 \pi}{\bt^*} , \al_i \right)$.
{\it In the present framework, the Gauss sum $\widehat{\la}_{\bt^*}$ powered by
$\left( 1 + \frac{i}{\im \tau} \right)$ is the only operator
that involves the odd prime number $\bt$.
Thus it is natural to interpret it as a creation operator of
the prime $\bt$.
Notice that, as discussed in (\ref{5-26}), the factor of
$\left( 1 + \frac{i}{\im \tau} \right)$ is necessary in
taking account of quantum realization of the Legendre symbols.}
This point becomes crucial in the following arguments.

Now it is well-known that Riemann's zeta function $\zeta (r)$ can be
expressed as a product sum over primes
\beq
    \zeta (r) \, = \, \prod_{P} \frac{1}{1 - P^{-r}}
    \, = \, \frac{1}{1 - 2^{-r}} \, \prod_{p} \frac{1}{ 1 - p^{-r}} \, .
    \label{5-38}
\eeq
This is known as Euler's product. Here
$P$ runs over all primes including $P=2$, while $p$ denotes odd primes as before.
The above analysis suggests that quantum theoretically the odd primes
$p$ can be replaced by the operator
$( \widehat{\la}_{p^*})^{1 + \frac{i}{\im \tau} }$, {\it i.e.},
\beq
    p \, \leftrightarrow \,
    ( \widehat{\la}_{p^*} )^{ 1 + \frac{i}{\im \tau} }
    = ( \sqrt{p} )^{ 1 + \frac{i}{\im \tau} } \, .
    \label{5-39}
\eeq
For $P=2$, either of the quantity $P^*$ or
the Gauss sum $\widehat{\la}_{2}$ is not defined.
However, one may define the Legendre symbol $\la_2 (x)$
where $x \in {\bf F}_{2}^{\times}$ takes a value of $x=1$ or $-1$.
In fact, it is possible to generalize the definition of
the Gauss sum such that it includes the case of $\la_2 (x)$.
This is a Gauss sum that is expanded by the so-called Dirichlet character $\chi (x)$
\cite{Dictionary}:
\beqar
    \widehat{\la}_P &=& \sum_{x \in {\bf F}_P} \chi (x) e^{i\frac{2 \pi}{P} x}
    \nonumber \\
    &=& \left\{
    \begin{array}{l}
    \sqrt{P} ~~~~ \mbox{for $\chi (-1) = 1$} \, ; \\
    i \sqrt{P} ~~~ \mbox{for $\chi (-1) = -1$} \, .  \\
    \end{array}
    \right.
    \label{5-40}
\eeqar
This can be seen as a generalization of (\ref{5-6}).
In this generalization, the Legendre symbol $\la_{P} (x)$ $(x \in {\bf F}_P )$
is interpreted as the Dirichlet character $\chi (x)$.
For $P=2$, we have $\chi (-1) = \la_{2} ( -1 ) = \left( \frac{-1}{2} \right)
= 1$. Thus the Gauss sum $\widehat{\la}_2$ can be calculated as
\beq
    \widehat{\la}_2 \, = \, \sqrt{2} \, .
    \label{5-41}
\eeq

Now we would like to define $2^*$. A naive application of the definition
$q^* = (-1)^{\frac{q-1}{2}}q$ leads to $2^{*} = i 2$.
However, we may consider that this is inappropriate because
$2^*$ should take a value of either $+2$ or $-2$ as the rest
of prime numbers. We then {\it define} $2^*$ as $2^*  =  2$ so that we also obtain
\beq
    \widehat{\la}_{2^*} \, = \, \sqrt{2} \, .
    \label{5-42}
\eeq
This is a natural choice if we think of the fact that
the elements of ${\bf F}_2$ and ${\bf F}_{-2}$
are identical; for both cases the elements can be given by either of $\pm 1$.
Of course, this fact itself does not justify the
choice of $2^* = 2$ but there is nothing
that prohibits this choice as well since the notion of $2^*$ is
introduced in a way that is somewhat independent of the
correspondence between knots and primes in (\ref{5-1}).
In fact, the essential quantities
in the following discussions are the moduli (or the absolute values)
of the Gauss sums.
Thus, although the distinction between $P$ and $P^*$
is important in relation between linking numbers and Legendre symbols,
it does not affect the modulus of the Gauss sum,
$| \widehat{\la}_{P} | =| \widehat{\la}_{P^*} | = \sqrt{P}$.
This indicates another justification of the choice of $2^* = 2$.
With (\ref{5-42}) understood along these lines of reasonings,
we find that the relation (\ref{5-39}) holds for any prime numbers $P$ in general.
Therefore the operator $\Theta_{R, \ga}^{(\tau)} ( a + 1 + i \im \tau ; \bt )$
in (\ref{5-37}) can further be written as
\beqar
    && \Theta_{R, \ga}^{(\tau)} ( a + 1 + i \im \tau ; \bt )
    \nonumber \\
    &=& \Tr_{R, \ga } \, \exp \Bigg[
    \sum_{r \ge 2} \, \bt^{ \frac{r}{2} \left( 1+ \frac{i}{\im \tau} \right) }
    \, \d_{a_1} \otimes \d_{a_2} \otimes \cdots \otimes \d_{a_r}
    \nonumber \\
    && ~~~~~~~~~~~~~~~~~~~~~~~~~~~~~~~~~~~~~ (-1)^{r}
    \, \prod_{P} \frac{1}{ 1 - {P}^{ \frac{-r}{2} \left( 1+ \frac{i}{\im \tau} \right) } }
    \Bigg]
    \label{5-43}
\eeqar
where $\bt$ is an odd prime number and $P$ runs over all prime numbers.
By use of this holonomy operator, we can calculate an analog of ``scattering amplitude''
for $n$ $\bt$'s:
\beqar
    \A ( \bt^n ) &=& \left. \frac{1}{ ( n-1 )! }
    \frac{\del}{\del \d_{\d_{a_1}}} \otimes
    \frac{\del}{\del \d_{\d_{a_2}}} \otimes \cdots
    \frac{\del}{\del \d_{\d_{a_n}} }
    \Theta_{R , \ga}^{(\tau)} ( a+ 1 + i \im \tau ; \bt )
    \right|_{\d_a = 0}
    \nonumber \\
    &=&
    (-1)^{n}
    \bt^{\frac{n}{2} \left( 1+ \frac{i}{\im \tau} \right)}
    \prod_{P} \frac{1}{1 - P^{\frac{-n}{2} \left( 1+ \frac{i}{\im \tau} \right)}}
    \nonumber \\
    &=&
    (-1)^{n}
    \bt^{\frac{n}{2} \left( 1+ \frac{i}{\im \tau} \right)}
    \, \zeta \left( \frac{n}{2} + i \frac{n}{2 \im \tau} \right)
    \label{5-44}
\eeqar
where $n \ge 2$.
In the holonomy formalism this can be considered as
a scattering amplitude of $n$ odd primes of the same value $\bt$.
From a perspective of particle physics, we can regard $\bt$ as a boson
of a same species. Thus (\ref{5-44}) can also be interpreted as
a nobel scattering amplitude of bosons.
The square of the amplitude is given by
\beq
    \left| \A ( \bt^n ) \right|^{2} \, = \, \bt^n \,
    \left| \zeta \left( \frac{n}{2} + i \frac{n}{2 \im \tau} \right) \right|^{2} \, .
    \label{5-45}
\eeq
The appearance of Riemann's zeta function
indicates that there is a nontrivial self-interactions
among $\bt$'s depending on the number of $\bt$'s that involves
in the scattering processes.
In the large $n$ limit, we have
\beq
    \left| \zeta \left( \frac{n}{2} + i \frac{n}{2 \im \tau} \right) \right|^{2}
    \, \rightarrow \, 1
    ~~~ ( n \gg 1 ) \, .
    \label{5-46}
\eeq
This suggests that the self-interaction effect vanishes for sufficiently large $n$.
This asymptotic behavior also supports the idea that the appearance of Riemann's
zeta function arises form a quantum effect of the scattering processes among
identical odd primes.

\vskip 0.5cm \noindent
\underline{Physical interpretation of the Riemann hypothesis}

Riemann's zeta function $\zeta (s)$ $(s \in {\bf C})$
vanishes at the negative even integers, {\it i.e.},
$s = -2, \, -4, \, -6, \cdots \, $. These are called the trivial zeros
of $\zeta (s)$. There is a famous conjecture about nontrivial zeros
of $\zeta (s)$, that is, the real part of any nontrivial zeros of
$\zeta (s)$ is equal to $\frac{1}{2}$.
This is called the Riemann hypothesis.
In the following, we propose a physical interpretation of
this hypothesis by extrapolating the result (\ref{5-44}) to
the case of $n= 1$.

The zero-mode holonomy operator (\ref{5-43}) and the abelian holonomy
operator (\ref{2-15}) in general are defined for $n \ge 2$.
As discussed below (\ref{2-16}), we can properly define the exponent of
the holonomy operators to be zero for $r=1$.
Following the expressions in (\ref{r4-8})-(\ref{r4-9}), let us denote this quantity
as
\beq
    \Path^{(\tau)} \oint_{\ga} \widetilde{A} \, := \, \d_{a_1} \oint_{\ga} \om^{(1)} \, .
    \label{5-47}
\eeq
Notice that the iterated-integral representation of Riemann's zeta function (\ref{r4-8})
is not defined for $r = 1$.
This problem can however be solved by taking the limit of $x \rightarrow 1$
in the following relation:
\beqar
    \Li_{1} (x) &=& \int_{0}^{x} \om^{(1)}  = \int_{0}^{x} \frac{dt }{1 -t}
    = - \log (1-x)
    \nonumber \\
    & \rightarrow & \oint_{\ga} \om^{(1)} = \zeta (1) ~~~~ (x \rightarrow 1)
    \label{5-48}
\eeqar
where $\ga$ is defined in (\ref{4-22}) and
$\om^{(1)}$ is defined in (\ref{r4-0}), respectively.
Thus, by use of the polylogarithm function, we can properly extrapolate
the expression (\ref{r4-8}) to the case of $r=1$.
As is well-known, the resultant value $\zeta (1)$ diverges.
This seems to cause a problem in defining the quantity (\ref{5-47}) being zero.
But this can easily be remedied by
assuming that the expectation value of $\d_{a_1}$ or $\frac{\pi}{\im \tau} \ba_1$
can be expanded by factors of $| 1 - x |^\ep$ with $\ep > 0$.
In principle, such an assumption can always be imposed as
there are no restrictions on the asymptotic behaviors of
the zero-mode variables at $x \rightarrow 1$.

The above argument shows that we can properly define the exponent of
the zero-mode holonomy operator even for $r=1$.
The gauged holonomy operator (\ref{5-43}) is obtained
from the application of the zero-mode holonomy operator (\ref{4-37}) to a space of
odd prime number $\bt$.
The structure of the gauged holonomy operator
(\ref{5-43}) and the origin of
Riemann's zeta function in particular remain the same as
the original holonomy operator (\ref{4-37}).
The difference arises from the introduction of the Gauss-sum
operator (powered by a certain factor) which we interpret
as a creation operator for the odd prime $\bt$.
We then further rewrite the zeta function
in terms of this prime-creation operator, using the formula of Euler's product.
Therefore we can similarly define
the case of $r=1$ for the gauged holonomy operator (\ref{5-43}) as well.
This means that the ``scattering amplitude'' in (\ref{5-44}) can be
applied for $n=1$, giving an expression
\beq
    \A ( \bt ) \,  =  \, -
    \bt^{ \left( \frac{1}{2} + i \frac{1}{2 \im \tau} \right)}
    \, \zeta \left( \frac{1}{2} + i \frac{1}{2 \im \tau} \right)
    \, = \, 0 \, .
    \label{5-49}
\eeq
This is nothing but the indication of the Riemann hypothesis in
the abelian holonomy formalism.

Physically it is obvious that the amplitude (\ref{5-49}) vanishes because
there is no notion of ``scattering'' for a single-particle system.
In other words, there is no notion of ``interaction'' in
a one-body system. This assertion is more likely
true for a system of a massless boson.
In this sense, the result (\ref{5-49}) can be physically well-understood.
Furthermore, in the present case, we no longer deal with the divergent
$\zeta (1)$ so that here we do not have to really worry about the
asymptotic behavior of the zero-mode variables which we have discussed above.

The imaginary part of the nontrivial zeros of $\zeta (s)$ are numerically given by
$14.13, ~ 21.02, ~ 25.01, \cdots$ for $\im s > 0$.
The largest value of $\im \tau$ that satisfies (\ref{5-49}) is then evaluated as
\beq
     ( \im \tau )_{\rm max} = 0.03537 \, .
     \label{5-50}
\eeq
This value decreases towards zero as $\im s$ increases.
Since $\tau = i \im \tau$ is the modular parameter of the torus,
the value of $\im \tau$ characterizes the shape of the torus, {\it i.e.},
the Riemann surface on which we define the two-dimensional conformal field theory.

%%%%%%%%%%%%%%%%%%%%%%%%%%%%%%%%%%%%%%%%%%%%%%%%%
\section{Concluding remarks}

In the present paper, we further investigate an abelian version of
the holonomy formalism that is recently developed
as a nonperturbative approach to non-abelian gauge theories \cite{Abe:2009kn}.
We first review the construction of a holonomy operator
in two-dimensional conformal field theory
and then consider zero-mode contributions to it.
The zero-mode analysis is made possible by use of a geometric-quantization scheme.
We then develop a method to extract a purely zero-mode part of the
abelian holonomy operator by redefining the ``path ordering'' in the operator.
It turns out that the zero-mode holonomy operator $\Theta_{R , \ga}^{(\tau )} ( a) $
can be expressed in terms of Riemann's zeta function.

Along the way, we also show that a generalization of linking numbers can be
obtained in terms of the vacuum expectation values of the zero-mode
holonomy operators.
This is shown by use of an abelian gauge theory on the zero-mode variables
$a_i \in {\bf C}$ ($i= 1,2, \cdots , n$) where
gauge transformations are realized by the double periodicity condition
$a_i \rightarrow a_i + m_i +  n_i  \tau$ ($m_i , n_i \in {\bf Z}$).
This means that $a_i$ is nothing but a complex coordinate
on a torus with a modular parameter $\tau$ which we set to
$\tau = i \im \tau$.
The integers $m_i$ and $n_i$ correspond to the winding numbers of
$\al_i$ and $\bt_i$ cycles for the $i$-th torus, respectively.
As discussed in the end of section 3, by fixing one of these numbers, we can
interpret the other as a linking number between the $\al_i$ and $\bt_i$ cycles.

Inspired by mathematical analogies between linking numbers
and Legendre symbols, we then consider an application
of the gauged zero-mode holonomy operator
$\Theta_{R , \ga}^{(\tau )} ( a + 1 + i \im \tau) $ to a space of
${\bf F}_{p}^{\times}$ where $p$ is an odd prime number.
This enables us to incorporate Legendre symbols into
$\Theta_{R , \ga}^{(\tau )} ( a + 1 + i \im \tau)$
such that they can be treated as operators in the space of ${\bf F}_{p}^{\times}$.
We make both classical and quantum analyses on this new
holonomy operator and show that it naturally leads to
the so-called Jacobi symbols by a suitable choice of $\al_i$ and $\bt_i$,
with $\bt_i$ being identical for any $i$'s.

Furthermore, we utilize the fact that the Gauss sum can be interpreted
as a Fourier transform (or a Fourier expansion, to be more precise)
of the Legendre symbol.
An analog of the phase space that is relevant to the Fourier transform
is given by $\left( \frac{2 \pi}{p} , x \right)$ with $x \in {\bf F}_{p}^{\times}$.
Thus, in the conjugate space of $\frac{2 \pi}{p}$, the Legendre symbol
can be represented by the Gauss sum.
We apply these mathematical facts to $\Theta_{R , \ga}^{(\tau )} ( a + 1 + i \im \tau)$
and construct a gauged zero-mode holonomy operator
$\Theta_{R , \ga}^{(\tau )} ( a + 1 + i \im \tau ; \bt )$
in a space that is conjugate to a space of ${\bf F}_{\bt^*}^{\times}$, with
$\bt^*$ defined by $\bt^* = (-1)^{\frac{\bt -1}{2}} \bt$.
An explicit form of the operator
$\Theta_{R , \ga}^{(\tau )} ( a + 1 + i \im \tau ; \bt )$
is shown in (\ref{5-43}).

What is interesting, if not mathematically rigorous, in obtaining
the expression (\ref{5-43}) is that we identify the Gauss-sum
operator $\widehat{\la}_{\bt^*}$ as a creation operator of an odd prime $\bt$.
As mentioned in section 5, in an operator language, the Gauss sum powered
by the factor of $\left( 1 + \frac{i}{ \im \tau } \right)$
is the only operator that involves the odd prime $\bt$.
Thus it is natural, at least from a perspective of physicists',
to interpret $\widehat{\la}_{\bt^*}$ as an operator that is relevant to
the creation of prime number $\bt$.
An explicit correspondence is given in (\ref{5-39}).
In section 5, we also discuss that this interpretation
can naturally be extended to the case of the even prime number, $\bt = 2$.
By use of this correspondence and the formula of Euler's product,
we can therefore make a persuasive argument to express
Riemann's zeta function in terms of the Gauss-sum operators.
One may understand this expression as a consequence of the application
of abelian holonomy formalism to the elementary theory of numbers.
In this sense, the expression $\Theta_{R , \ga}^{(\tau )} ( a + 1 + i \im \tau ; \bt )$
in (\ref{5-43}) presents a quantum realization
of Riemann's zeta function in the holonomy formalism.

% say about RH
The holonomy formalism is a kind of universal formalism that uses a holonomy operator
of conformal field theory as an S-matrix functional for scattering amplitudes
of massless bosons.
In this context, we can identify the gauged zero-mode holonomy operator
$\Theta_{R , \ga}^{(\tau )} ( a + 1 + i \im \tau ; \bt )$
as a generating function for an analog of ``scattering amplitude''
for identical prime numbers.
The resultant ``scattering amplitude'' is then computed in (\ref{5-44}).
We argue that this result can also be applied to the case of
a single-particle, or a single-prime, system and shows that
in this case the result (\ref{5-44}) provides a novel indication
of the Riemann hypothesis. Physically this result arises from an obvious
fact that there is no notion of interaction in a single-particle system.

% relation to primon gas system
The reader may wonder whether or not our formulation has any relevance to
the so-called primon gas system, also known as the free Riemannian gas system,
in statistical physics \cite{Julia:1990}.
In the primon gas system, one considers a set of
non-interacting bosonic particles (or primons) each of which
is labeled by a prime $p_i$, having an eigenvalue $E_i = E_0 \log (p_i )$.
The suffix $i$ here is put to label a distinct prime.
Using the prime factorization of an integer $N$ (see (\ref{5-8})
for its concrete form), we can uniquely characterize
a multi-particle system by $N$.
The partition function of the primon gas system then becomes
$Z = \sum_{N \ge 1} N^{- E_0 / k_B T} = \zeta ( E_{0} / k_{B} T )$
where $k_B$ is Boltzmann's constant and $T$ denotes the temperature of the system.
This is a basic result of the primon gas system and illustrates an interesting
connection between number theory and physics.
(For other known examples of such connections, focusing
on physical interpretations of the zeta function and particularly the Riemann
hypothesis, see a recent review \cite{Schumayer:2011yp}.)
What is intriguing in the present paper, in this context, is that
we indicate an underlying conformal field theory for the interactions
among many primons, inspired by Morishita's analogy between primes and knots
\cite{Morishita:2009gt}.
Thus our formulation is qualitatively different form the primon gas system;
we are not considering {\it free} primons
but somewhat {\it interacting} primons in the calculation of physical quantities.
For example, we show that unnormalized ``scattering probabilities'' of odd primes
involving a specific prime factorization can be calculated as $\zeta ( n )^2$
where $n$ denotes the number of distinct primes
appearing in the prime factorization; see equations (\ref{5-34}).
This result provides a new physical interpretation of
Riemann's zeta function. It also illustrates that our formulation may well
open up a new concept of interacting, rather than non-interacting, primons.

Lastly, we would like to remark that our construction of
zero-mode holonomy operators provides an interesting framework
for studies on quantum aspects of topology and number theory.
The gauged zero-mode holonomy operators in forms of
(\ref{5-28}) and (\ref{5-43}) define abelian gauge theories on zero modes.
These are simple $U(1)$ gauge theories but, as we have
seen explicitly, corresponding physical results (\ref{5-34})
and (\ref{5-45}) hold regardless the choices of
$\al_i$ and $\bt$ which correspond to either an odd prime number
or an integer.
The choices depend on how we realize the correspondence between
linking numbers and Legendre symbols.
Furthermore, as shown in (\ref{5-33}), for a certain choice
we do not have to fix the winding numbers $m_i$ and $n_i$,
or at least one of them, to the identity.
Thus the abelian gauge theories which we construct
have indeed very rich properties in terms of topology
and number theory, and will be useful for studies of
these subjects in a framework of quantum field theory.

%%%%%%%%%%%%%%%%%%%%%%%%%%%%%%%%%%%%%%%%%%%%%%%%%
\vskip .4in
%\newpage
\section*{Appendix: Index of notations}
%\begin{tabular}{l l}
\begin{longtable}{l l}
  %\hline
  $a_i ~  ( \rightarrow \, a_i + m_i + n_i  \tau ) $
  & complex variables for zero modes ($m_{i}, n_{i} \in {\bf Z}$) \\
  $a_i^{(\pm)}$ & creation operators of photons with helicity $\pm$  \\
  & corresponding to the ladder operators of $SL(2,{\bf C})$\\
  $A = \frac{1}{\kappa} \sum_{i < j} A_{ij} \, \om_{ij}$ & comprehensive gauge fields for photons \\
  $A_{ij} = a_{i}^{(+)} \otimes a_{j}^{(0)} + a_{i}^{(-)} \otimes a_{j}^{(0)}$ & bialgebraic operators \\
  $A^{(\tau )} = \sum_{i < j} A^{(\tau)}_{ij} \, \om_{ij} $ & comprehensive gauge fields for zero modes \\
  $A^{(\tau)}_{ij} =  1_i \otimes \d_{a_j}$ & bialgebraic operators for zero modes\\
  $\B_n = \Pi_{1} ( \C )$ & braid group \\
  $\C = {\bf C}^n / {\S_n}$ & physical configuration space \\
  $\eta_{\mbox{{\tiny $N$}}} (y) = \left( \frac{y}{N} \right)$ & Jacobi symbol ($y \in {\bf Z}$) \\
  ${\bf F}_p = {\bf Z}/p {\bf Z}$ & finite field (reduced residue class group) \\
  ${\bf F}_{p}^{\times} = {\bf F}_p - \{ 0 \}$ & space of mod $p$\\
  $i$, $j$ $(= 1,2, \cdots , n)$ & numbering index \\
  $k$ $(=1)$ & level number of the $U(1)$ Chern-Simons theory \\
  $\kappa $ & KZ parameter \\
  $K ( a_i , \ba_i )$ & K\"{a}hler potential for zero modes \\
  $\la_p (q) = \left( \frac{q}{p} \right)$ & Legendre symbol \\
  $\widehat{\la}_p = \sum_{x \in {\bf F}_{p}^{\times} } \la_p (x) \exp \left( \frac{i 2\pi}{p} x \right)$
  & Gauss sum, equal to $\sqrt{p^*}$ \\
  $m_i$, $n_i$  ($=1$)& winding numbers of $\al_i$ and $\bt_i$ cycles, respectively \\
  $N = p_{1}^{e_1} p_{2}^{e_2} \cdots p_{s}^{e_s}$ & prime factorization of odd positive integer $N$ \\
  $\om_{ij} = d \log (z_i - z_j )$ & logarithmic one-form  \\
  $\om^{(0)}_{j} = \frac{d z_j}{ z_j }$ & logarithmic one-form (type 0) \\
  $\om^{(1)}_{j} = \frac{ - d z_j }{1 - z_j}$ & logarithmic one-form (type 1)\\
  $\Om = \frac{1}{\kappa} \sum_{ i < j } \Om_{ij} \om_{ij}$  & KZ connection \\
  $\Om_{ij} = \sum_{\mu} a_{i}^{(\mu )} \otimes a_{j}^{(\mu )}$ & bialgebraic operators,
  $\mu (= 0, 1, 2)$ corresponding \\
  & to the full indices for the elements of $SL (2 ,{\bf C})$ \\
  $\Om^{(\tau )}$ & zero-mode K\"{a}hler form on torus \\
  $p$, $q$ & odd prime numbers \\
  $p^* = (-1)^{\frac{p-1}{2}} p$ &  odd primes with sign $\pm$ \\
  $P$ & prime numbers (including $P=2$) \\
  $\S_n$ & rank-$n$ symmetric group \\
  $\tau = \re \tau + i \im \tau = i \im \tau$ & modular parameter of torus, setting $\re \tau = 0$ \\
  $\Theta_{R , \ga} (z) $ & abelian holonomy operator of $A$ \\
  $\Theta_{R , \ga} (z , a) $ & abelian holonomy operator of $\widetilde{A} = A + A^{(\tau )}$ \\
  $\Theta_{R , \ga}^{(\tau )} ( a) $ & zero-mode holonomy operator; $A^{(\tau )}$-part of
  $\Theta_{R , \ga} (z , a)$ \\
  $\Theta_{R , \ga}^{(\tau )} ( a + 1 + i \im \tau) $ & Gauged zero-mode holonomy operator \\
  $\Theta_{R , \ga}^{(\tau )} ( a + 1 + i \im \tau ; \bt , \al_i ) $ &
  $\Theta_{R , \ga}^{(\tau )} ( a + 1 + i \im \tau) $ incorporated with $\la_{\bt^*} ( \al_i )$ \\
  & where $\bt$ is an odd prime and $\al_i \in {\bf F}_{\bt^*}^{\times}$ \\
  $\Theta_{R , \ga}^{(\tau )} ( a + 1 + i \im \tau ; \bt ) $ &
  $\Theta_{R , \ga}^{(\tau )} ( a + 1 + i \im \tau) $ incorporated with $\widehat{\la}_{\bt^*}$ \\
  $V^{\otimes n} = V_1 \otimes V_2 \otimes \cdots \otimes V_n$ & physical Hilbert space \\
  $V_i$ & Fock space for the $i$-th photon \\
  $\zeta (r) = \sum_{n \ge 1} \frac{1}{n^r}$ & Riemann's zeta function \\
  %\hline
%\end{tabular}
\end{longtable}

%%%%%%%%%%%%%%%%%%%%%%%%%%%%%%%%%%%%
%\vskip .3in
%\noindent
%{\bf Acknowledgments} \vskip .06in\noindent
%The author thanks the Yukawa Institute for Theoretical Physics at Kyoto University.
%Discussions during the YITP workshop YITP-W-08-04 on
%``Development of Quantum Field Theory and String Theory'' were useful for the present work.

\vskip .3in
%%%%%%%%%%%%%%%%%%%%%%%%%%%%%%%%%%%%%%%%%%%%%%%%%%%%%%%%%%%%%%%%


\begin{thebibliography}{99}


%%% Topology and CFT %%%
%\cite{Kohno:2002bk)
\bibitem{Kohno:2002bk}
  T.~Kohno, {\it Conformal Field Theory and Topology}, Translations of
  Mathematical Monographs, Volume 210, American Mathematical Society (2002).
%\cite{Witten:1988hf}
\bibitem{Witten:1988hf}
  E.~Witten,
  %``Quantum field theory and the Jones polynomial,''
  Commun.\ Math.\ Phys.\  {\bf 121}, 351 (1989).
  %%CITATION = CMPHA,121,351;%%

%%% Holonomy formalism  %%%
%\cite{Abe:2009kn}
\bibitem{Abe:2009kn}
  Y.~Abe,
  %``Holonomies of gauge fields in twistor space 1: bialgebra, supersymmetry,
  %and gluon amplitudes,''
  Nucl.\ Phys.\  B {\bf 825}, 242 (2010)
  [arXiv:0906.2524 [hep-th]].
  %%CITATION = NUPHA,B825,242;%%

%%% Linking number and abelian CS theory %%%
%\cite{Polychronakos:1989-90}
\bibitem{Polychronakos:1989-90}
  A.~P.~Polychronakos,
  Annals Phys.\  {\bf 203}, 231 (1990); Phys.\ Lett.\  B {\bf 241}, 37 (1990).
%\cite{Kaul:1999je}
\bibitem{Kaul:1999je}
  R.~K.~Kaul,
  ``Topological quantum field theories: A meeting ground for physicists and  mathematicians,''
  arXiv:hep-th/9907119.
  %%CITATION = HEP-TH/9907119;%%
%\cite{Guadagnini:2010gy}
\bibitem{Guadagnini:2010gy}
  E.~Guadagnini,
  ``Functional integration and abelian link invariants,''
  arXiv:1001.4645 [hep-th].
  %%CITATION = ARXIV:1001.4645;%%

%%% Algebraic topology; degrees of mapping %%%
%\cite{Fulton:1995bk}
\bibitem{Fulton:1995bk}
  W.~Fulton, {\it Algebraic Topology: A First Course}, Springer (1995).

%%% Abelian CS on torus: our approach %%%
%\cite{BosNair:1989}
\bibitem{BosNair:1989}
  M.~Bos and V.~P.~Nair,
  Phys.\ Lett.\  B {\bf 223}, 61 (1989); Int.\ J.\ Mod.\ Phys.\  A {\bf 5}, 959 (1990).
%\cite{NairBook}
\bibitem{NairBook}
  V.~P.~Nair, {\it Quantum Field Theory: A Modern Perspective},
  Springer (2004), see pp.515-522.
%\cite{Abe:2008wn}
\bibitem{Abe:2008wn}
  Y.~Abe,
  %``On the deconfining limit in (2+1)-dimensional Yang-Mills theory,''
  Nucl.\ Phys.\  B {\bf 828}, 215 (2010)
  [arXiv:0804.3125 [hep-th]].
  %%CITATION = NUPHA,B828,215;%%

%%% Linking number; various approaches %%%
%\cite{Horowitz:1989km}
\bibitem{Horowitz:1989km}
  G.~T.~Horowitz and M.~Srednicki,
  %``A Quantum Field Theoretic Description of Linking Numbers and Their Generalization,''
  Commun.\ Math.\ Phys.\  {\bf 130}, 83 (1990).
  %%CITATION = CMPHA,130,83;%%
%\cite{Ashtekar:1996bs}
\bibitem{Ashtekar:1996bs}
  A.~Ashtekar and A.~Corichi,
  %``Photon inner product and the Gauss-linking number,''
  Class.\ Quant.\ Grav.\  {\bf 14}, A43 (1997)
  [arXiv:gr-qc/9608017].
  %%CITATION = CQGRD,14,A43;%%
%\cite{Adams:1996yf}
\bibitem{Adams:1996yf}
  D.~H.~Adams,
  ``R-torsion and linking numbers from simplicial abelian gauge theories,''
  arXiv:hep-th/9612009.
  %%CITATION = HEP-TH/9612009;%%
%\cite{GarciaCompean:2009zh}
\bibitem{GarciaCompean:2009zh}
  H.~Garcia-Compean and R.~Santos-Silva,
  %``Link Invariants for Flows in Higher Dimensions,''
  J.\ Math.\ Phys.\  {\bf 51}, 063506 (2010)
  [arXiv:0908.3218 [hep-th]].
  %%CITATION = ARXIV:0908.3218;%%
  
%%% Primes and Knots; algebraic number theory %%%
%\cite{Morishita:2009gt}
\bibitem{Morishita:2009gt}
  M.~Morishita,
  ``Analogies between Knots and Primes, 3-Manifolds and Number Rings,''
  arXiv:0904.3399 [math.GT]; {\it Knots and Primes} (in Japanese), Springer-Japan (2009).
%\cite{Ono:1987bk}
\bibitem{Ono:1987bk}
  T.~Ono, {\it An Introduction to Algebraic Number Theory} (in Japanese),
  Shokabo (1987); Plenum Press (1990), English translation.
%\cite{Kohno:2006bk}
\bibitem{Kohno:2006bk}
  T.~Kohno and M.~Morishita (editors), {\it Primes and Knots},
  Contemporary Mathematics, Volume 416, American Mathematical Society (2006).

%%% Application of the iterative solution to a physical system
%\cite{Efraty:1992gk}
\bibitem{Efraty:1992gk}
  R.~Efraty and V.~P.~Nair,
  %``Action for the hot gluon plasma based on the Chern-Simons theory,''
  Phys.\ Rev.\ Lett.\  {\bf 68}, 2891 (1992)
  [arXiv:hep-th/9201058].
  %%CITATION = PRLTA,68,2891;%%
%\cite{Zupnik:1987vm}
\bibitem{Zupnik:1987vm}
  B.~M.~Zupnik,
  %``THE ACTION OF THE SUPERSYMMETRIC N=2 GAUGE THEORY IN HARMONIC SUPERSPACE,''
  Phys.\ Lett.\  B {\bf 183}, 175 (1987).
  %%CITATION = PHLTA,B183,175;%%

%%% Riemann zeros, physical interpretations %%%
%\cite{Sierra:2008se}
\bibitem{Sierra:2008se}
  G.~Sierra and P.~K.~Townsend,
  %``Landau levels and Riemann zeros,''
  Phys.\ Rev.\ Lett.\  {\bf 101}, 110201 (2008)
  [arXiv:0805.4079 [math-ph]].
  %%CITATION = PRLTA,101,110201;%%
%\cite{He:2009yp}
\bibitem{He:2009yp}
  Y.~H.~He, V.~Jejjala and D.~Minic,
  ``Eigenvalue Density, Li's Positivity, and the Critical Strip,''
  arXiv:0903.4321 [math-ph]; ``On the Physics of the Riemann Zeros,''
  arXiv:1004.1172 [hep-th].
  %%CITATION = ARXIV:0903.4321;%%
%\cite{Giri:2009pk}
\bibitem{Giri:2009pk}
  P.~R.~Giri,
  ``Physical realization for Riemann zeros from black hole physics,''
  arXiv:0905.4939 [hep-th].
  %%CITATION = ARXIV:0905.4939;%%

%%% books %%%
%\cite{Chari:1994pz}
\bibitem{Chari:1994pz}
  V.~Chari and A.~Pressley,
  {\it A Guide To Quantum Groups},
  %\href{/spires/find/hep/www?irn=3125289}{SPIRES entry}
  Cambridge University Press (1995).

%\cite{Dictionary}
\bibitem{Dictionary}
  Mathematical Society of Japan, {\it Sugaku Jiten} (in Japanese),
  Iwanami Shoten (1985); {\it Encyclopedic Dictionary
  of Mathematics}, MIT press (1987), English translation.

%\cite{Julia:1990}
\bibitem{Julia:1990}
  Bernard L. Julia,
  ``Statistical theory of numbers'' in
  {\it Number Theory and Physics}, edited by
  M. Waldschmidt, J. M. Luck and P. Moussa (Springer, Berlin) 1990, pp. 276-293.

%\cite{Schumayer:2011yp}
\bibitem{Schumayer:2011yp}
  D.~Schumayer and D.~A.~W.~Hutchinson,
  %``Physics of the Riemann Hypothesis,''
  Rev.\ Mod.\ Phys.\  {\bf 83}, 307 (2011)
  [arXiv:1101.3116 [math-ph]].
  %%CITATION = ARXIV:1101.3116;%%

\end{thebibliography}
\end{document}